\documentclass[twocol]{ametsocV6.1}

\newcommand{\re}{\mathrm{re}}
\newcommand{\im}{\mathrm{im}}

\newcommand\abs[1]{\lvert#1\rvert}

\title{The QBO, the annual cycle, and their interactions: Isolating periodic modes with Koopman analysis}

\authors{Claire Valva,  \aff{a}\correspondingauthor{Claire Valva, clairev@nyu.edu}
Edwin P. Gerber \aff{a}}

\affiliation{\aff{a}{Center for Atmosphere Ocean Science, Courant Institute of Mathematical Sciences, New York University, New York, New York}}

\abstract{The Quasi-Biennial Oscillation (QBO) is the dominant mode of variability in the equatorial stratosphere. It is characterized by alternating descending easterly and westerly jets over a period of approximately 28 months. It has long been known that the QBO interactions with the annual cycle, e.g., through variation in tropical upwelling, leading to variations in the descent rate of the jets and, resultingly, the QBO period. Understanding these interactions, however, has been hindered by the fact that conventional measures of the QBO convolve these interactions. Koopman formalism, derived from dynamical systems, allows one to decompose spatio-temporal datasets (or nonlinear systems) into spatial modes that evolve coherently with distinct frequencies. We use a data-driven approximation of the Koopman operator on zonal-mean zonal-wind to find modes that correspond to the annual cycle, the QBO, and the nonlinear interactions between the two. From these modes, we establish a data-driven index for a ``pure’' QBO that is independent of the annual cycle and investigate how the annual cycle modulates the QBO. We begin with what is already known, quantifying the Holton-Tan effect, a nonlinear interaction between the QBO and the annual cycle of the polar stratospheric vortex. We then use the pure QBO to do something new, quantifying how the annual cycle changes the descent rate of the QBO, revealing annual variations with amplitudes comparable to the $30 \, \mathrm{m} \, \mathrm{day}^{-1}$ mean descent rate. We compare these results to the annual variation in tropical upwelling and interpret them with a simple model.}

\begin{document} 

%% Necessary!
\maketitle

%%%%%%%%%%%%%%%%%%%%%%%%%%%%%%%%%%%%%%%%%%%%%%%%%%%%%%%%%%%%%%%%%%%%%
% SIGNIFICANCE STATEMENT/CAPSULE SUMMARY
%%%%%%%%%%%%%%%%%%%%%%%%%%%%%%%%%%%%%%%%%%%%%%%%%%%%%%%%%%%%%%%%%%%%%
%
% Significance Statement (all journals except BAMS)
%
\statement
The Quasi-Biennial Oscillation (QBO) is a periodic cycle of winds in tropical atmosphere with a period of 28 months. The phase of QBO is known to influence other aspects of the atmosphere, including the polar vortex, but the magnitude of its effects and how it behaves are known to depend on the season. In this study, we use a data-driven method (called a Koopman decomposition) to quantify annual changes in the QBO and investigate their causes. We show that seasonal variations in the stratospheric upwelling play an important but incomplete role.

%%%%%%%%%%%%%%%%%%%%%%%%%%%%%%%%%%%%%%%%%%%%%%%%%%%%%%%%%%%%%%%%%%%%%
% MAIN BODY OF PAPER
%%%%%%%%%%%%%%%%%%%%%%%%%%%%%%%%%%%%%%%%%%%%%%%%%%%%%%%%%%%%%%%%%%%%%
%
\section{Introduction} The Quasi-Biennial Oscillation (QBO) is the leading mode of variability of the equatorial stratosphere. It is characterized by downward propagating easterly and westerly wind regimes with a period of approximately 28 months \citep{baldwinQuasibiennialOscillation2001}. While a tropical phenomenon, the QBO is known to affect other regions of the atmosphere such as extratropical surface variability \citep{garfinkelInfluenceQuasibiennialOscillation2010, ansteyHighlatitudeInfluenceQuasibiennial2014}. The most well known teleconnection is the Holton-Tan effect: a warming of the boreal polar stratosphere during easterly QBO \citep{holtonQuasiBiennialOscillationNorthern1982}.  

While the oscillations of the QBO are perhaps the most regular mode in the climate system that are not directly linked to the diurnal or seasonal cycles, the period of the QBO ranges from 24 to 34 months. A key mechanism for the range of periods is the interaction between the QBO and the annual cycle \citep{hampsonPhaseAlignmentTropical2004,krismerSeasonalAspectsQuasibiennial2013,rajendranSynchronisationEquatorialQBO2016}. As the mean QBO period is not an integer multiple of the annual cycle, any interactions between the two occur irregularly. (The same could be said for interactions between the QBO and other regular oscillations in the climate system, e.g., the El Ni\~{n}o Southern Oscillation or the 11-year solar cycle.) A consequence of this irregular interaction means that it is hard to quantify the effects of the annual cycle on the QBO, or to establish what the QBO would look like during a given period in the absence of the annual cycle. 

Commonly used indices of the QBO, such as the zonal-mean zonal-wind at one (or more) pressure levels, or representations in EOF space, e.g., Wallace et al. (1993), include nonlinear modulations of the QBO by the annual cycle.  This occurs even when the input data is deasonalized before the analysis. Removing the seasonal cycle eliminates the linear, or average, influence of annual variations across all phases of the QBO, but does not account for variations that depend on the QBO phase.  An example of such a “nonlinear interaction” between frequencies is the Holton-Tan effect: planetary waves propagate deeper into the tropics when it is \emph{both} boreal winter \emph{and} the QBO is in a westerly phase.  A more sophisticated data analysis technique is required to account for nonlinear interaction between frequencies.

%For example, commonly used indices of the QBO — including the value of zonal-mean zonal-wind at one (or more) pressure levels or representations in EOF space, e.g., \citet{wallaceRepresentationEquatorialStratospheric1993} — will necessarily include nonlinear modulations of the QBO phase by the annual cycle. While the use of deseasonalized data to study the QBO can remove mean annual effects, the resulting index will still retain nonlinear interactions between the annual cycle and the QBO without a more complex data analysis technique to account for them.

In this study, we propose the use of Koopman methods to help solve this problem. We use a Koopman decomposition to categorize and separate nonlinear interactions of the QBO and the annual cycle, which allows for improved quantitative understanding of the QBO in the context of its irregular interaction with the annual cycle. Koopman operator theory translates between finite-dimensional nonlinear dynamical systems and linear (albeit infinite-dimensional) systems. As such, data-driven approximations of Koopman operators can provide decompositions of the climate system without assuming linearity. These decompositions identify a collection of quasi-periodic modes and corresponding frequencies from which one can analyze interactions between two modes of known frequencies.

The separation between the known frequencies of the annual cycle and the QBO allow Koopman methods to isolate the phenomena and analyze the interaction between them separately. An immediate result of this property is that we can create an objective QBO index (and corresponding resulting ``pure'' QBO Koopman mode) that is independent of the annual cycle. 

\begin{figure}[h]
    \centering
    \includegraphics[width=0.5\textwidth]{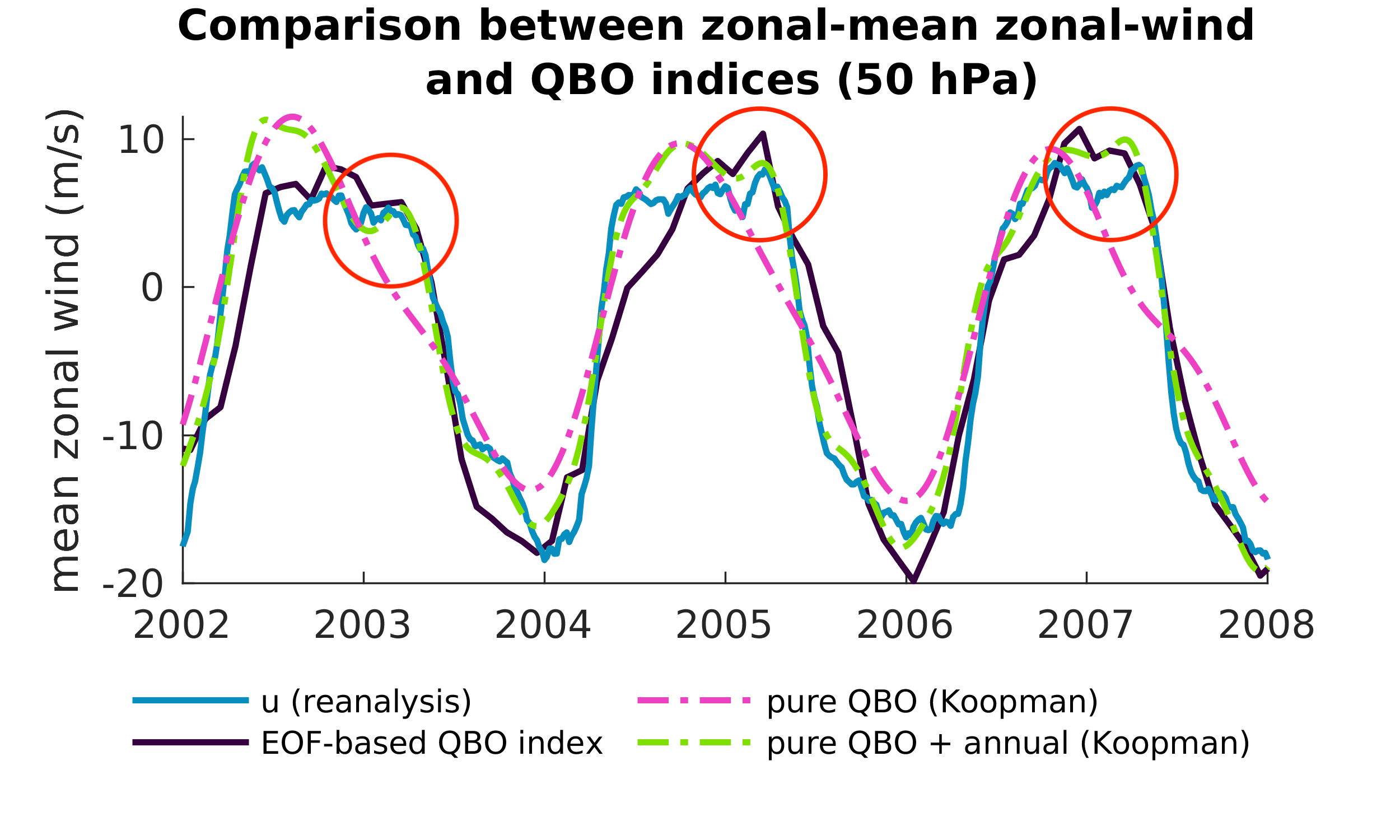}
    \caption{Comparison between raw ERA5 reanalysis zonal-mean zonal-wind, averaged from $-10^\circ$ to $10^\circ$ N (blue), an EOF based QBO index (dark purple), a pure QBO mode (Koopman derived, pink), and a combination QBO and annual cycle mode (Koopman derived, green) at 50 hPa. Reanalysis and the Koopman derived modes are mean zonal-mean zonal-wind between . The Koopman derived modes have been normalized to have the same mean as the reanalysis winds. Similarly, the EOF index has been rescaled to have the same standard deviation as the Koopman derived mode (pure QBO with annual interactions) and the mean of the reanalysis for comparative purposes. A ``shoulder'' in the zonal-wind when the westerly QBO phase occurs during early boreal winter is indicated by red circles.}
    \label{fig:index}
\end{figure}

To motivate the necessary mathematics, we entice readers with the potential of the method. Figure \ref{fig:index} compares indices from two Koopman derived QBO modes to raw ERA5 reanalysis zonal-wind \citep{hersbachERA5GlobalReanalysis2020} and an empirical orthogonal function (EOF) QBO index derived from Singapore radiosonde monthly mean winds (from \citet{newmanEOFs1980presentQBO2023}, computed as in \citet{wallaceRepresentationEquatorialStratospheric1993}), all at $50 \, \mathrm{hPa}$. Each index tracks the oscillation of the zonal-mean zonal-wind at $50 \, \mathrm{hPa}$ fairly similarly, switching from easterly and westerly jets around the same time. The pure QBO Koopman mode more smoothly tracks this evolution without impact from the annual cycle and higher frequency variability. The EOF based index, as well as our ``combination'' Koopman mode (which is explicitly designed to include nonlinear interactions between the QBO and annual cycle) retain interannual variability. In particular one observes a shifting ``shoulder'' (highlighted with red circles) when the westerly phase of the QBO falls early in the calendar year. This feature is a result of the interaction between the QBO and the annual cycle, which will be more illuminatingly discussed in Section \ref{sec:construction}\ref{sec:modes}. While the conventional EOF based QBO index was computed from deseasonalized data, it clearly retains seasonal effects, more closely following the Koopman mode built to include them.  

The remainder of the manuscript is organized as follows. In Section \ref{sec:koopman} we give a brief overview of Koopman theory and our data-driven algorithm, followed by a description of how to create and interpret aggregate Koopman modes in Section \ref{sec:construction}, where the term \textit{aggregate} refers to the inclusion of harmonic frequencies which gives a more complete representation. We use these aggregate Koopman modes to analyze the Holton-Tan effect and variation of QBO descent rates with the annual cycle in Section \ref{sec:results}. Finally, in Section \ref{sec:discussion} we discuss other possibilities for use of this method, including application to other phenomena and model assessment.

\section{Koopman methods: Theory and application}
\label{sec:koopman}
We begin with a brief explanation of Koopman operator theory and our data-driven Koopman approximation. We use the computational method for approximating the Koopman operator developed in \citet{dasReproducingKernelHilbert2021}. For a complete overview of this method, as well as precise mathematical statements of Koopman operators, we refer the reader to the appendices and supplementary material of \citet{froylandSpectralAnalysisClimate2021} and \citet{lintnerIdentificationMaddenJulian2023} for a detailed description of the algorithm and to \citet{dasReproducingKernelHilbert2021} for the development of the algorithm and relevant spectral convergence results. The following discussion, however, should be sufficient for understanding the results and conclusions of this study.

\subsection{Koopman formalism}
\label{sec:theory}
Consider a dynamical system with states denoted $\{x_t\}_{t \in \mathbb{R}}$ which is associated with a continuous map $\varphi^s$ that takes the state at time $t$ and pushes it forward to time $s + t$, i.e., $\varphi^s(x_t) = x_{s + t}$. As a parallel, consider a numerical model. We may use a numerical model to approximate $\varphi^s$, integrating the dynamical system $\frac{d}{dt}x = f(x)$ to determine $x_{s + t}$ based on the initial condition $x_t$.

Associated with dynamical systems are scalar functions $g$ called \textit{observables} that take the state space as their domain. For example, a possible $g$ for the state space of the climate system would be the function that computes or ``observes'' the mean temperature of the atmosphere. Then $g(x_t)$ would be the mean temperature at time $t$. 

The \textit{Koopman operator} $K^s$ describes the evolution of the state of a dynamical system $\{x_t\}_{t \in \mathbb{R}}$ by the action on observables of the system:
\begin{equation}
    (K^s g)(x_t) = g \circ \varphi^s(x_t)  = g(x_{s + t}).
\end{equation}
In our example, $K^sg$ maps the observation of mean atmosphere temperature forward by time $s$. The operator $K^s$ is linear on the space of observables (or functions of the state) even when the dynamics $\varphi$ are nonlinear. For a wide class of systems of interest — i.e., those with an invariant probability measure (see \citet{dasReproducingKernelHilbert2021} for a precise statement) — there is an alternative formulation of $K^s$ by a skew-adjoint linear operator $V$, which is called the \textit{Koopman generator}, that is analogous to a derivative.
\begin{subequations}
\begin{align}
    Vg = \lim_{t \to 0} \frac{K^t g - g}{t} \label{eq:twoa} \\
    K^t g = e^{tV} g, \label{eq:twob}
\end{align}
\end{subequations}
where the exponential of an operator is defined using the series form for $e$, i.e., $e^A = \sum_k \frac{1}{k!}A^k$.Here, eq. (\ref{eq:twoa}) defines the action of $V$ onto the function $g$, and eq. (\ref{eq:twob}) shows how $V$ generates the Koopman operator $K^t$. 

This formulation allows a spectral decomposition of $V$ to serve as coherent feature extraction for dynamical systems through the eigenproblem:16
\begin{equation}
    e^{tV} \zeta_k = e^{t\omega_k} \zeta_k,
\end{equation}
where the eigenfunctions $\zeta_k$ evolve predictably according to the eigenfrequency $\omega_k$. As $V$ is a skew-adjoint operator in this setting (where there is a invariant probability measure), all $\omega_k$ will be purely imaginary. As such, the eigenfunctions $\zeta_k$ are periodic modes that capture the evolution operator. If the evolution of the full system (or a particular feature of interest) is well described by a reasonably small set of eigenfunctions $\zeta_k$, then there is an efficient representation of the dynamics that is inherently predictable.

The Koopman formalism also allows for eigenfrequency-eigenfunction generation given already known eigenfunctions. Suppose that you have eigenfrequencies $\alpha$ and $\beta$ which correspond to the eigenfunctions $a$ and $b$. Then there will also be an eigenpair with eigenfrequency $i(\alpha + \beta)$ and eigenfunction $ab$ \citep{reedMethodsModernMathematical1972},
\[ e^{Vt}a b = e^{i\alpha t} a e^{i\beta t} b = e^{i(\alpha + \beta)t} ab. \]
We discuss the relevant consequences of this construction property in Section \ref{sec:construction}\ref{sec:eigenfreqs}, but one can think of this generation property as accounting for harmonics (of the form $2\alpha$ and $a^2$) and nonlinear interactions between components of a dynamical system with different frequencies (here, $\alpha$ and $\beta$). In later sections, we focus on the nonlinear interactions of this kind that can be physically interpreted, i.e., that have eigenfrequencies of the form $\alpha + \beta$ where $\alpha$ and $\beta$ correspond to a frequencies of interest. Other nonlinear interactions may exist in the data that cannot be identified in this way.

\subsection{Data-driven Koopman approximation}
\label{sec:datadriven}
We can find decompositions of the climate system without assuming linearity by implementing this Koopman formalism. Given data, we want to find the Koopman eigenfunctions $\zeta_k$ and associated frequencies $\omega_k$ to build a decomposition of Koopman modes.

This method provides a dynamically meaningful decomposition that can represent nonlinear dynamics, overcoming some of the limitations of methods such as EOFs \citep{monahanEmpiricalOrthogonalFunctions2009}. Other methods for extracting (and prediction of) oscillations from the climate system include singular spectrum analysis \citep{ghilAdvancedSpectralMethods2002a} and linear inverse models \citep{penlandStochasticModelIndoPacific1996, albersSubseasonalPredictabilityNorth2021}. We are interested in studying \textit{nonlinear interactions}, and the Koopman approximation method chosen here is skilled at finding nonlinear interactions via eigenfrequency generation, which we discuss more in section \ref{sec:construction}, as well as having convergence and stability guarantees in the large data limit. Previous uses of this method have included analysis of the El Ni\~no Southern Oscillation (ENSO, \citet{froylandSpectralAnalysisClimate2021}) and identification of the Madden-Julian Oscillation (MJO, \citet{lintnerIdentificationMaddenJulian2023}).

\begin{figure}[h]
    \centering
    \includegraphics[width=0.5\textwidth]{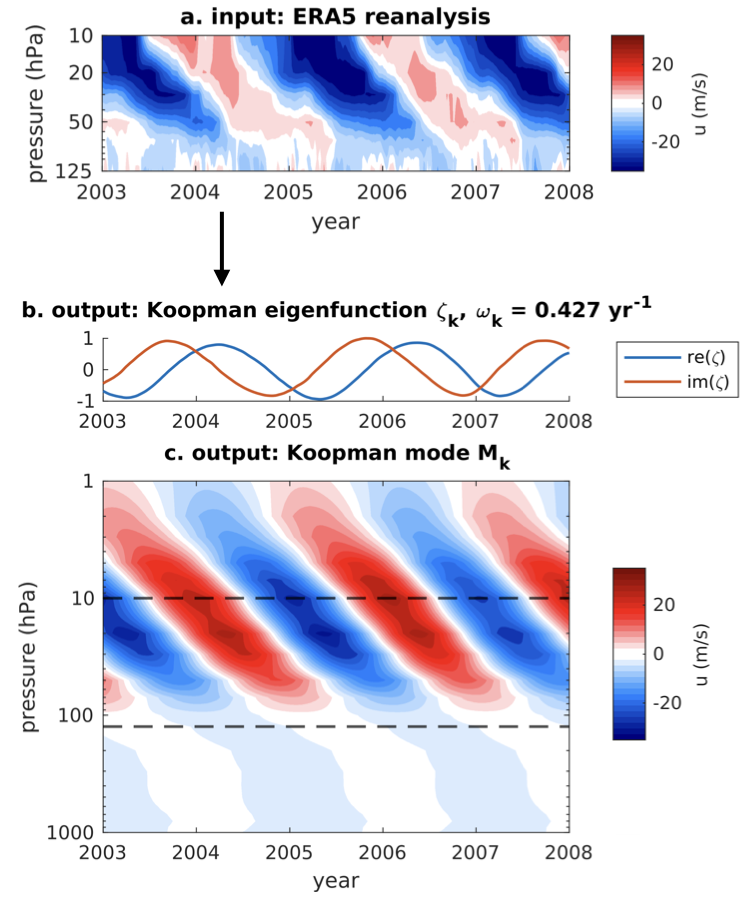}
    \caption{Visual overview of Koopman decomposition method that shows the input data (a) zonal-mean zonal-wind averaged between $-10^\circ$ and $10^\circ$ N, (b) one of the $N$ scalar Koopman eigenfunctions, here with frequency $\omega = 0.42737 \, \mathrm{yr}^{-1}$ and (c) zonal-mean, zonal-wind between $-10^\circ$ and $10^\circ$ N of the associated mode $M_k$, which is computed by projecting reanalysis data onto the eigenfunction. Note that the y-axis range changes between top and bottom panels. The dashed lines in (c) correspond to the range of the data in (a).}
    \label{fig:scheme}
\end{figure}

From given input data $\mathcal{D}$ of size $N_t \times N_d$ (time by space), we approximate the Koopman generator $V$ with $N$ eigenfrequency-eigenfunction pairs $(\omega_k, \zeta_k)$ of $V$. As $V$ is a skew-adjoint operator, eigenvalues should be purely imaginary and come in complex conjugate pairs, and we index our eigendecomposition such that $\im(\omega_k) = -\im(\omega_{-k})$ and $\zeta_k = \bar{\zeta}_{-k}$. Each eigenfunction $\zeta_k$ will have mean zero. Diffusion is introduced to ensure computational stability, resulting in the eigenvalues having a negative real part. As such, with this particular algorithm, values of $\abs{\re(\omega_k)}$ increase monotonically with the Dirichlet energy ($E_k = \int |\nabla \zeta_k |^2$), a measure of eigenfunction regularity. As such, we order the pairs $(\omega_k, \zeta_k)$ such that $\abs{\re(\omega_k)}$ increases with $k$. This ordering serves as a proxy for numerical approximation error and has been used with empirical success \citep{dasReproducingKernelHilbert2021}.

The Koopman eigenpair corresponding to an extremely regular cycle (such as the annual or diurnal cycles) will be easier to resolve numerically due to its regularity. As the magnitude $\re(\omega_k)$ increases with the Dirichlet energy (and as such, eigenfunction regularity), we can alternatively interpret the magnitude of $\re(\omega_k)$ to roughly correspond to the periodicity of a given Koopman mode is: the smaller $\abs{\re(\omega_k)}$, the more periodic. There exists an alternative eigenvalue ordering based on autocorrelation decay (proposed in \citet{giannakisConsistentSpectralApproximation2024}) which gives similar results.

Additionally, there will always be an eigenfrequency $\omega_0 = 0$, where $\zeta_0 \equiv 1$ is constant to which corresponds to the mean state of a given system. We then have $N$ Koopman eigenpairs where $N = 2L + 1$, for $L$ complex conjugate pairs of frequencies $\omega_k$ and the a single mean state corresponding to $\omega_0$. 

From these eigenpairs, we compute \textit{Koopman modes} $M_k$ from linear projections of wanted target data onto the eigenfunction $\zeta_k$. As we obtain $M_k$ via projection, $M_k$ is not limited to the size or variables of the initial input data $\mathcal{D}$ and can instead be extended to a larger physical area, different observed variable, or extended time period (similarly to EOFs). An important difference in output between the Koopman analysis and EOFs, however, is that the modes $M_k$ are not necessarily orthogonal to each other. If we were to compute $M_k$ via a projection on the initial input data for all $k$ we can recover the complete data, i.e., $\sum^{N}_k M_k = \mathcal{D}$ when the number of modes $N$ is equal to the number of time samples $N_t$. However, we choose to compute only a limited number of modes $N \ll N_t$ for better computational and analysis efficiency: for very large $k$, we expect the modes $M_k$ to be less well-resolved, less approximately periodic, and to be of generally small magnitude. The goal is to find a small subset of well-resolved modes that capture the variability of interest.

In our study, we project the Koopman eigenfunctions onto data on all pressure levels while the input data for the eigenproblem is only levels between the upper troposphere to the middle stratosphere. In figure \ref{fig:scheme}, we give a visual overview of the inputs and outputs of this method. 
We show a sample interval of the input data (zonal-mean zonal-wind from 10 to 125 $\mathrm{hPa}$) and one of the output Koopman eigenpairs. The Koopman eigenfunction $\zeta_k$ is associated with eigenfrequency $0.42737 \, \mathrm{yr}^{-1} \approx 28$ month period, and the Koopman mode $M_k$ is given from the projection of the zonal-mean zonal-wind from 1 to 1000 $\mathrm{hPa}$ onto $\zeta_k$. This Koopman mode has many aspects of the QBO, as to be expected from its period, including the downward propagating easterly and westerly jets. As discussed in section \ref{sec:construction}, a more complete representation of the QBO requires inclusion of additional harmonics. 

One potential  limitation of the used method is the assumption that the dynamical system is stationary with an invariant probability measure. Clearly, this assumption is not strictly true for post-industrial climate data, but this violation does not appear to greatly affect results. Highly non-stationary data will result in a decomposition with eigenfunctions with nonzero mean and a purely real eigenfunction --- essentially trend modes in the decomposition. No trends appear in our analysis of the historical record, but they could (and do) appear in analysis of  longer climate change projection simulations or variables with more obvious trends, such as tropical sea surface temperatures.

Finally, it has been found that approximations of this kind can be improved with a technique called \textit{delay embedding}, e.g., \citet{ghilAdvancedSpectralMethods2002a, giannakisDatadrivenSpectralDecomposition2019} Here, we replace the input data $\mathcal{D}$ with a larger $\hat{\mathcal{D}}$ of size $(N_t - (N_e - 1)) \times (N_d \cdot N_e)$. Now each row of $\hat{\mathcal{D}}$ includes observations from the $N_e - 1$ previous time points, i.e., 
\begin{equation}
    \hat{\mathcal{D}}_t = (\mathcal{D}_t, \mathcal{D}_{t - 1}, \dots , \mathcal{D}_{t - (N_e - 1)}).
    \label{eq:delay}
\end{equation}
This acts to insert memory of previous time steps into the data. The computation of Koopman eigenvalues and eigenfunctions proceeds as previously described, where we substitute $\hat{\mathcal{D}}$ for $\mathcal{D}$. A rule-of-thumb often used to choose the number of delays $N_e$ to be on the order of the timescale of the phenomena of interest. For example, in this work studying the QBO, we use an embedding corresponding to approximately 23 months. 

\section{Construction of a pure QBO and related modes via Koopman decomposition}
\label{sec:construction}
We perform and interpret a Koopman decomposition of zonal-mean zonal-wind of the stratosphere. We use these results to create the QBO index (seen in figure \ref{fig:index}), as well as aggregate Koopman modes that correspond to the construction of a pure QBO mode, an annual cycle mode, and a QBO-annual cycle interaction mode.

We use five-day averages of zonal-mean zonal-wind ERA5 reanalysis \citep{hersbachERA5GlobalReanalysis2020} from 1979 to 2020 at seven pressure levels between $125$ and $10$ hPa and include all latitudes. We choose a five-day average to decrease the high frequency variability and the computational expense. The data is density and area-weighted and then delay embedded with $N_e = 140$ samples. Results were similar, both qualitatively and quantitatively,  with varied embedding lengths and  a seven-day average rather than five. We compute 50 Koopman eigenfunctions (51 including the eigenpair corresponding to Koopman mode that is the time mean). The Koopman modes $M_k$ are projected onto ERA5 data at 21 pressure levels between $1000$ and $1$ hPa. 

\subsection{Interpretation of eigenfrequencies}
\label{sec:eigenfreqs}
\begin{figure}[h!]
    \centering
    \includegraphics[width=0.46\textwidth]{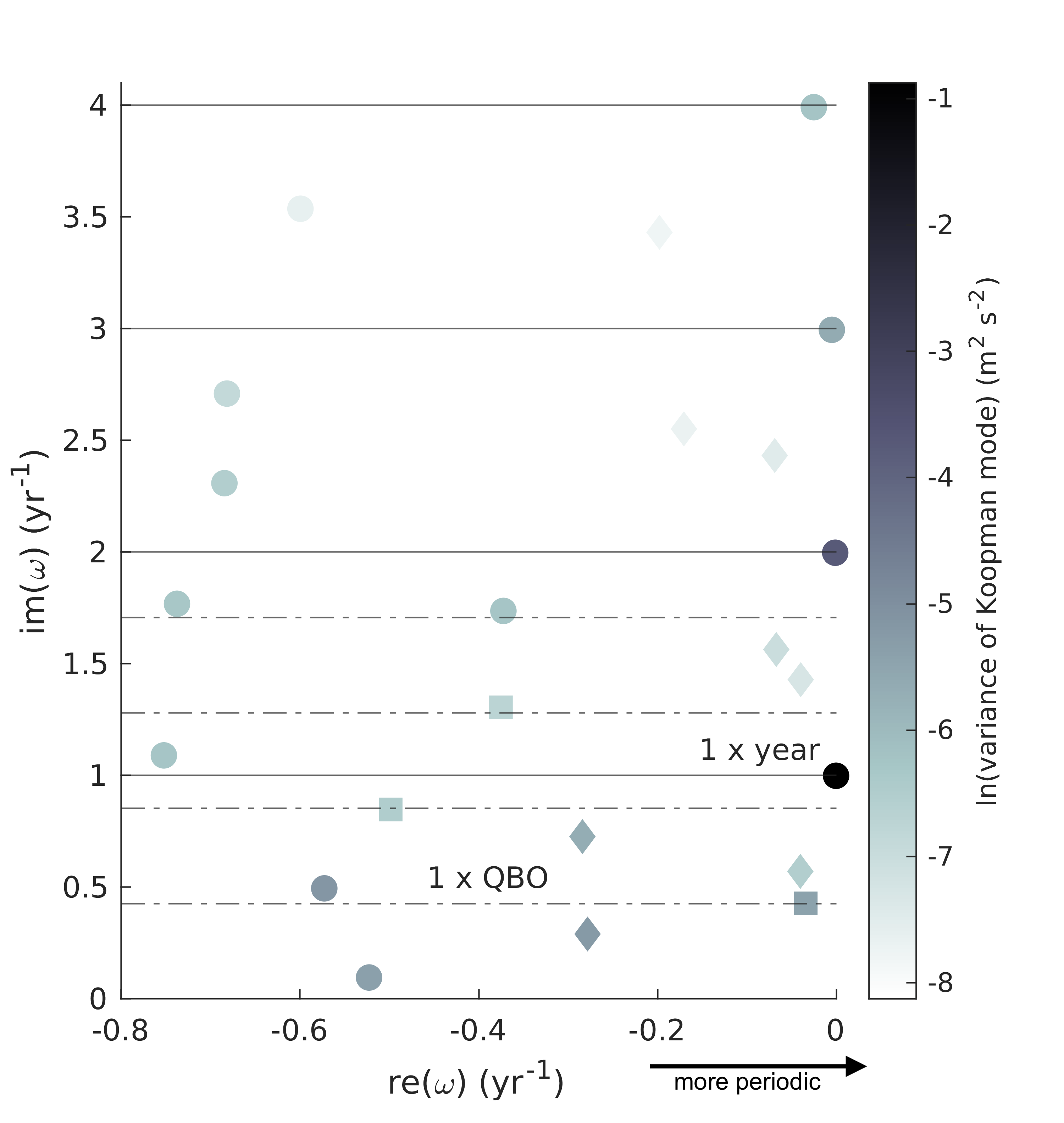}
    \caption{Numerically computed Koopman eigenvalues plotted on the complex plane. The y-axis value is the frequency of each eigenmode, while the x-axis can be used as a proxy for the quality or the periodicity of the numerical mode (as a perfect computation would have $\re(\omega)$ be 0). Color denotes the log of the density and area weighted variance of each Koopman mode, while dotted and solid horizontal lines denote frequencies corresponding to integer multiples of the QBO ($\sim 0.4286\:\mathrm{yr^{-1}}$, denoted with square markers) and the seasonal cycles respectively. Diamond shaped markers denote QBO-annual cycle interaction frequencies, i.e., those of the form $n \pm 0.4286 \: m$ within a 10-day tolerance on the desired period.}
    \label{fig:eigenvalues}
\end{figure}

Each Koopman mode has an associated frequency. For the well-resolved modes we want to be able to interpret these frequencies in terms of known phenomena. We emphasize that this method does not allow for targeting frequencies \textit{a priori}, i.e., there is no point in which we specify that we would like to find an eigenfrequency of once every 28 months or of the annual cycle. The input data can, of course, affect results: the delay embedding was chosen to emphasize interannual variability. Additionally, if the data is too coarsely sampled, we can miss oscillations. In this case, the diurnal cycle cannot be observed given that our input data has a sampling frequency of $5 \, \mathrm{days}^{-1}$. The pressure levels were chosen to prioritize variability at the levels of the QBO. Using input data that extended to $1$ hPa decreased resolution of the results, given the prevalence of the semiannual oscillation. 

In figure \ref{fig:eigenvalues}, we plot the all eigenvalues with positive frequencies (the eigenvalues of negative frequencies are simply the complex conjugates) in the complex plane, shaded by the log of the (density and area weighted) variance of each mode. As the eigenfunctions are ordered by their effective periodicity and regularity, one is not guaranteed that top the modes capture a substantive fraction of the variance. The Koopman eigendecomposition is constructive, meaning that we expect harmonics of significant modes (modes with integer multiples of the base frequency) as well as interactions between modes (modes that are integer linear combinations of frequencies). 

The leading eigenvalues — in both the sense of smallest real part (rightmost) and highest variance (darkest) — correspond to the annual cycle or harmonics thereof with integer valued frequencies. We also see frequencies that correspond to harmonics of the QBO, $0.4286 \, m \, \mathrm{yr}^{-1}$ for $m = 1, 2, \dots$, which correspond to a period of 28 months. Finally, we have identified modes that we call ``QBO-annual cycle interaction'' modes with frequencies that are approximately equal to $n + 0.4286\, m \, \mathrm{yr}^{-1}$ for integers $m$ and $n$ within a selected tolerance, chosen here to have a period that is within 10 days (twice the sampling frequency) of the desired period. 

Some eigenfrequencies remain unidentified. Of these, we conjecture that some could be sorted into the three previously discussed categories — particularly higher harmonics of the QBO or the interaction terms — but they do not fall into the desired frequency tolerance. Neglecting these modes minimally affects the results in later sections, as the unidentified modes are relatively small when compared to zonal-mean zonal-wind target data: the unidentified mode with the highest variance $(0.05 \, \mathrm{m}^2 \, \mathrm{s}^{-2})$ has a root-mean-square amplitude (over all levels) of $0.17 \,\mathrm{m} \, \mathrm{s}^{-1}$ and a maximum value (at any level) of $2.6\, \mathrm{m } \, \mathrm{s}^{-1}$. For reference, the zonal-mean zonal-wind target data has a variance of $327.0 \, \mathrm{m}^2 \, \mathrm{s}^{-2}$ and root-mean-square amplitude of $12.9 \, \mathrm{m } \, \mathrm{s}^{-1}$.

\subsection{Aggregate Koopman modes}
\label{sec:modes}
\begin{figure*}
    \centering
    \includegraphics[width=0.9\textwidth]{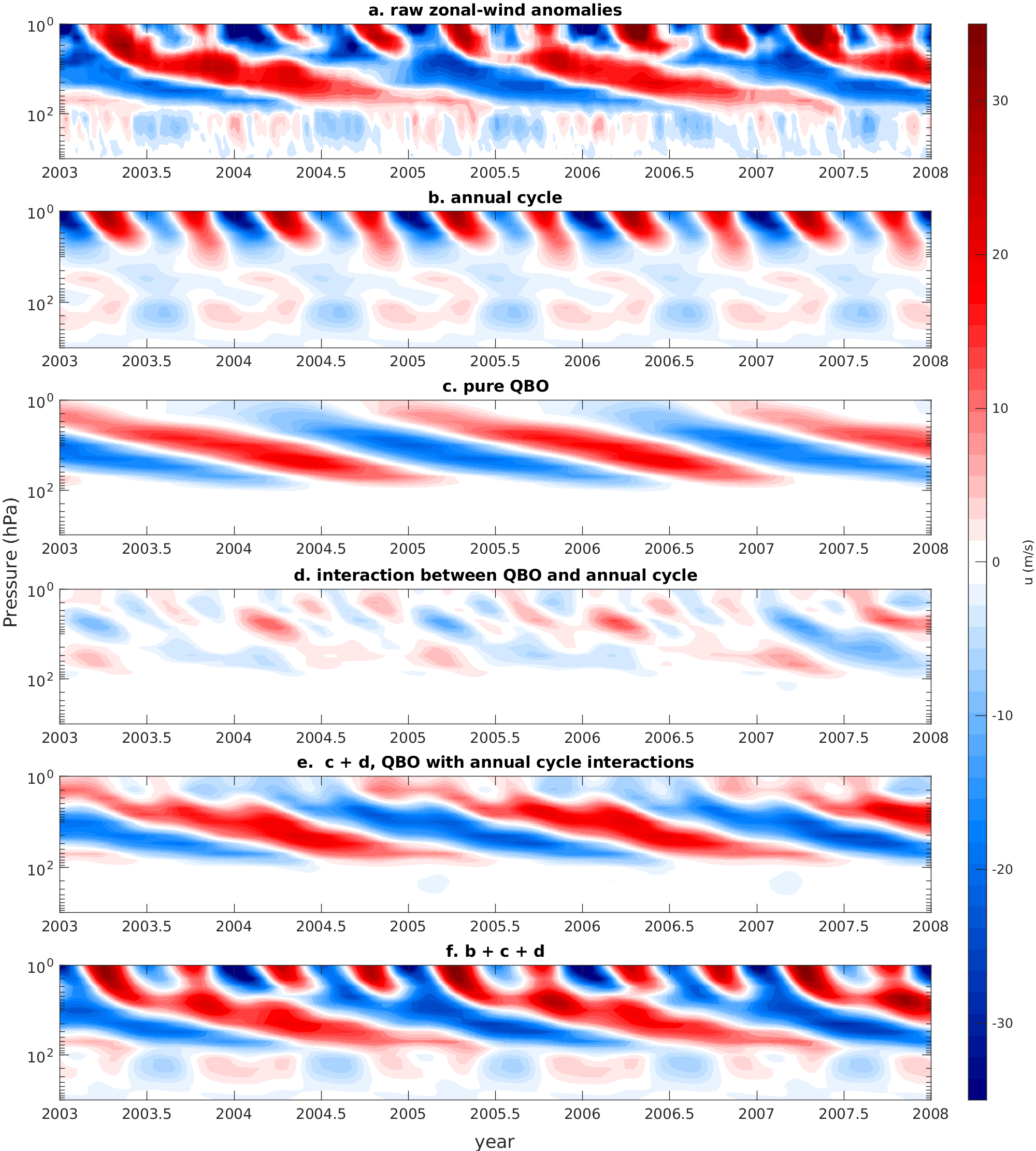}
    \caption{Comparison between the tropical mean zonal-wind and the aggregate modes representing the annual, pure QBO, and interaction between the QBO and the annual cycle. We plot means of zonal-mean zonal-wind between $-10^\circ$ and $10^\circ$ N of each component between 2003 and 2008. The time mean winds are removed for clarity. a: the raw wind anomalies. b: seasonal modes, i.e., approximately integer frequencies. c: pure QBO modes, frequencies that are multiples of $0.4286$. d: seasonal-QBO interaction modes, frequencies of the form $n \pm 0.4286 \, m$, e: total QBO, sum of panels (c) and (d), f: total seasonal and QBO effects, sum of panel (b) and (e). The final panel (f) reconstructs the total wind (a), except for short time variability.}
    \label{fig:initcomp}
\end{figure*}

From the eigenfrequency groupings discussed in Section \ref{sec:construction}\ref{sec:eigenfreqs}, we can create corresponding aggregate Koopman modes which we will refer to as an annual cycle mode (made up of integer frequencies), a pure QBO mode (frequencies of the form $ 0.4286 \, m$ for integers $m$), and an annual cycle-QBO interaction mode (frequencies of the form $n \pm 0.4286 \, m$). The annual cycle aggregate mode is the sum the projection of 5 eigenfunctions (and their conjugates), the pure QBO aggregate mode is made up of the projections of 3 eigenfunctions, and the annual cycle-QBO interaction aggregate mode is made up of 8.  Each of these aggregate modes are the sum of the Koopman modes $M_k$ that correspond to the desired $\omega_k$. For example, the aggregate mode for the annual cycle can be written as $M_{\mathrm{annual}} = \sum_{k, \, \omega_k \approx n} M_k$.  

The aggregate Koopman modes isolate particular phenomena: the pure QBO mode represents the evolution of the QBO in the absence of any interaction or interference by the seasonal cycle. The mean of the pure QBO mode in any month of the year, say December (not shown), is zero up to sampling uncertainty. Any impact of the annual cycle on the QBO will appear in the annual cycle-QBO interaction aggregate mode. 

Figure \ref{fig:initcomp} illustrates these aggregate modes and compares these to a slice of raw zonal-mean zonal-wind reanalysis (panel a). The sum of three aggregate modes (panel f) nearly captures the full reanalysis zonal-wind, but for high frequency variability, and shows how the flow is dominated by the QBO and annual cycle.  At first glance, the pure QBO mode (panel c) seems to lack some common characteristics of the QBO including the varying descent of the zero-line. The sum of the QBO and its interactions with the annual cycle (panel e) captures the variability of its descent rate. When the annual cycle is added (panel f), the bulk characteristics of the flow are recovered.

%However, the lack of these characteristics in the pure QBO mode can either be attributed to them being included in the mean state (the asymmetry between phases) or to nonlinear interaction between the QBO and a phenomenon with a different frequency (in the case of the zero-line descent rate, the annual cycle). 

\subsection{Koopman eigenfunction as QBO index}
\begin{figure*}
    \centering
    \includegraphics[width=\textwidth]{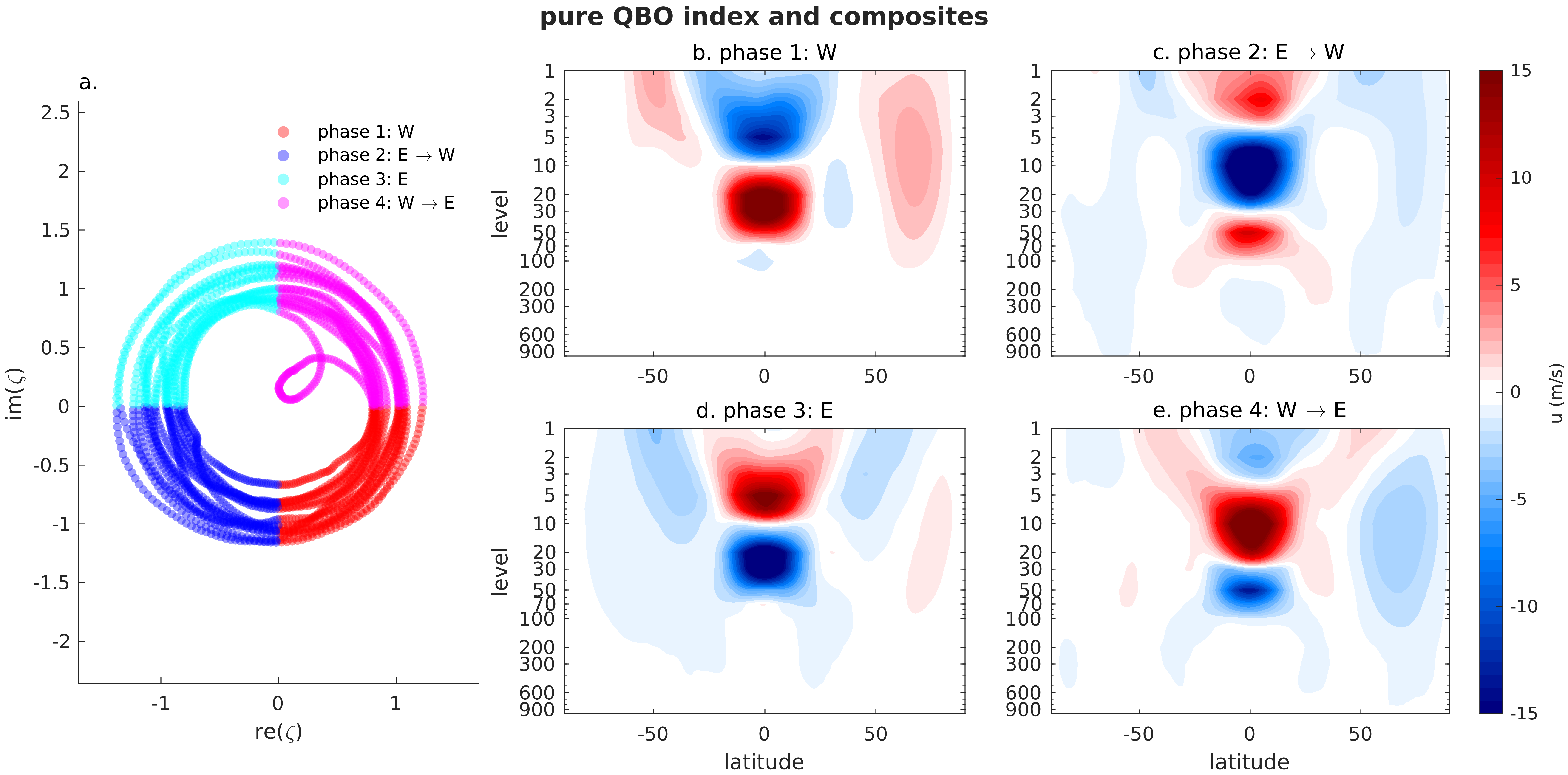}
    \caption{
        The scalar (complex) Koopman eigenfunction corresponding to eigenfrequency $\omega=0.427$ is used to make composites of the Koopman modes, achieved by splitting the QBO into four phases based on the quadrant of the Koopman eigenfunction. Colors of the scatter plot (panel a) of the Koopman eigenfunction correspond to phases one through four: a pure westerly phase ($1: W$), a descending easterly phase ($2: E \to W$), a pure easterly phase ($3: E $), and a descending westerly phase ($4: W \to E$). The ``loop'' in the pink scatter points (phase 4, descending westerly) corresponds to the disruption of the QBO that took place in 2016.}
    \label{fig:phases}
\end{figure*}
In addition to constructing aggregate Koopman modes, we can also use the Koopman eigenfunctions to create indices. Namely, we use this to create an ``objective QBO index'' from the single Koopman eigenfunction with frequency $\omega = 0.4274$ that is independent of annual effects. As we have a scalar eigenfunction, we can create a continuous index of the QBO defined by its phase angle. We plot this in figure \ref{fig:phases} and use this index to divide the QBO into 4 phases: a pure westerly phase (denoted as $1: W$), a descending easterly phase ($2: E \to W$), a pure easterly phase ($3: E $), and a descending westerly phase ($4: W \to E$). The number of phases is arbitrary, but four is chosen to balance a clarity of the QBO state and the need for sufficient data for statistics. 

The timespan plotted includes the QBO disruption that took place in 2016 that interrupted the descent of the westerly phase of the QBO \citep{bartonOrigin2016QBO2017}. It materialized in this QBO index as the diversion and loop from the usual oscillation that returned to its usual path after the conclusion of the disruption. 

Figure \ref{fig:phases} examines whether our pure QBO aggregate mode captures known characteristics of the QBO. While the QBO index is defined from a single Koopman eigenfunction, we plot composites of the pure QBO aggregate mode in the figure to include variability associated with harmonics. The composites capture the arch, or horseshoe-shaped, wind anomalies that extend downward from the QBO region in the subtropics, conjectured to be associated with the QBO mean meridional circulation by \citet{garfinkelInfluenceQuasiBiennialOscillation2011}. These can be identified in the phase 2, 3, and 4 composites as a downward extension of wind anomalies at beginning at approximately 50 hPa. Interestingly, the pure QBO mode captures modest strengthening (weakening) of the polar vortex during westerly QBO (easterly QBO), a shadow of the Holton-Tan effect. We investigate this particular aspect more in Section \ref{sec:results}\ref{sec:holtontan}.

\section{Understanding QBO-annual cycle interaction with Koopman modes}
\label{sec:results}
We can now use the aggregate Koopman modes to investigate the nonlinear interaction between the QBO and the annual cycle. We first replicate known results regarding the Holton-Tan effect (Section \ref{sec:results}\ref{sec:holtontan}) as further proof of concept for the use of these modes. We then use them to quantify the effect of seasonality on the QBO descent rate via comparison to the pure QBO mode (Section \ref{sec:results}\ref{sec:descentrate}).

\subsection{The Holton-Tan effect}
\label{sec:holtontan}
The Holton-Tan effect, first noted in \citet{holtonQuasiBiennialOscillationNorthern1982}, refers to the QBO influence on the strength of the wintertime stratospheric polar vortex in the Northern Hemisphere. Specifically, the polar vortex is weaker and warmer during easterly QBO (EQBO) and stronger and colder during westerly QBO (WQBO), as westerlies in the subtropics favor more equatorward propagation of planetary Rossby waves, thereby shielding the polar vortex \citep{luRoleRossbyWave2020}. 

\begin{figure*}
    \noindent\includegraphics[width=0.9\textwidth]{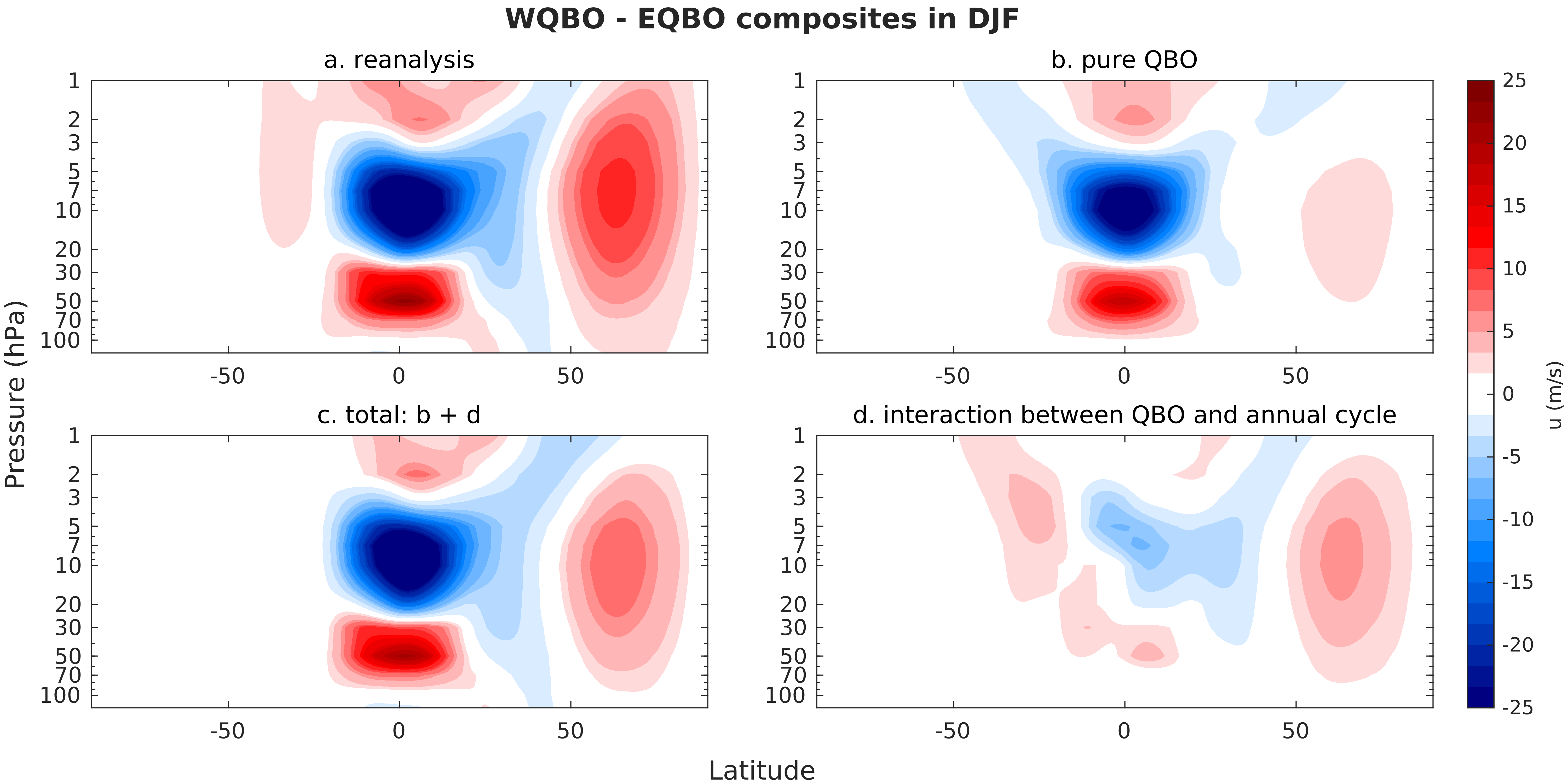}
    \caption{Composites of zonal-mean zonal-wind during boreal winter (DJF) of WQBO-EQBO as measured by mean winds over the season at 50 hPa. Panel (a) is reanalysis — the typical composite shown for the Holton-Tan effect. Panel (b) and panel (d) are compositions of the pure QBO aggregate mode and QBO-annual cycle interaction aggregate mode. Panel (c) is the sum of (b) and (d). In the case of a perfect reconstruction from the Koopman modes, we would expect (c) to be the same as (a); Differences between these panels are on the order of 3 $\mathrm{m}\mathrm{s}^{-1}$.}
    \label{fig:HT}
\end{figure*}

In figure \ref{fig:HT}, we show composites of EQBO subtracted from WQBO in boreal winter (DJF) using the full reanalysis record of winds, versus the winds captured by our aggregate Koopman modes. Panel (a) is a composite formed from ERA5 reanalysis, where climatology EQBO (WQBO) is defined as mean reanalysis winds at $50\, \mathrm{hPa}$ over the season being less than $-3 \, \mathrm{m} \, \mathrm{s}^{-1}$ (greater than $3 \mathrm{m} \,\mathrm{s}^{-1}$) and the two are then subtracted. The remaining three panels are formed similarly, where we compute climatologies for a given aggregate mode over the EQBO and WQBO defined from reanalysis.  We do not include the 2016 QBO disruption in these composite. 

Use of the Koopman modes allows us to separate the Holton-Tan effect into a pure QBO component that is independent of the time of year (figure \ref{fig:HT}b is identical for any other season; not shown) and a nonlinear interaction component that is only present during DJF (figure \ref{fig:HT}d). Figure \ref{fig:HT}b and the QBO index composites (figure \ref{fig:phases}b, e) indicate a mean strengthening (weakening) of the polar vortex during WQBO (EQBO). 

The annual cycle-interaction aggregate mode captures the strengthening of the polar vortex in DJF. This strengthening is reversed in other seasons, where the aggregate mode is weakly negative. This cancels out the mean strengthening of the polar vortex in the pure QBO composites in other seasons.

One could create an equivalent figure where the determination of whether a given winter belongs to either EQBO or WQBO categories is done by using either the phase of the scalar QBO index or value of the zonal-wind from the pure QBO aggregate mode. While either of these would be a more ``objective'' way to define the phases of the QBO, we kept with the standard convention for consistency. The other approaches produce similar results; see Appendix \ref{appendix:b}. Additionally, all the composites closely match similar composites in the literature \citep{baldwinQuasibiennialOscillation2001, luRoleRossbyWave2020}. 
% Similarly, an alternative tolerance of $\mp 5\, \mathrm{m} \,\mathrm{s}^{-1}$ produces minor changes in results.

The biggest differences between the Koopman reconstruction of the composite from reanalysis are present in the upper stratosphere, which indicates that we are excluding some stratospheric processes with our Koopman modes. We conjecture that this discrepancy may be partially due to the 11-year solar cycle, which has been suggested to influence the polar vortex and has a lower frequency than any of our resolved Koopman modes. In addition, this analysis does not account for ENSO interactions by design, which has also been noted to affect the polar vortex \citep{ansteyHighlatitudeInfluenceQuasibiennial2014}.

% previous citation works for both

\subsection{The annual cycle in QBO descent rate}
\label{sec:descentrate}
\begin{figure}[h]
    \centering
    \includegraphics[width=0.5\textwidth]{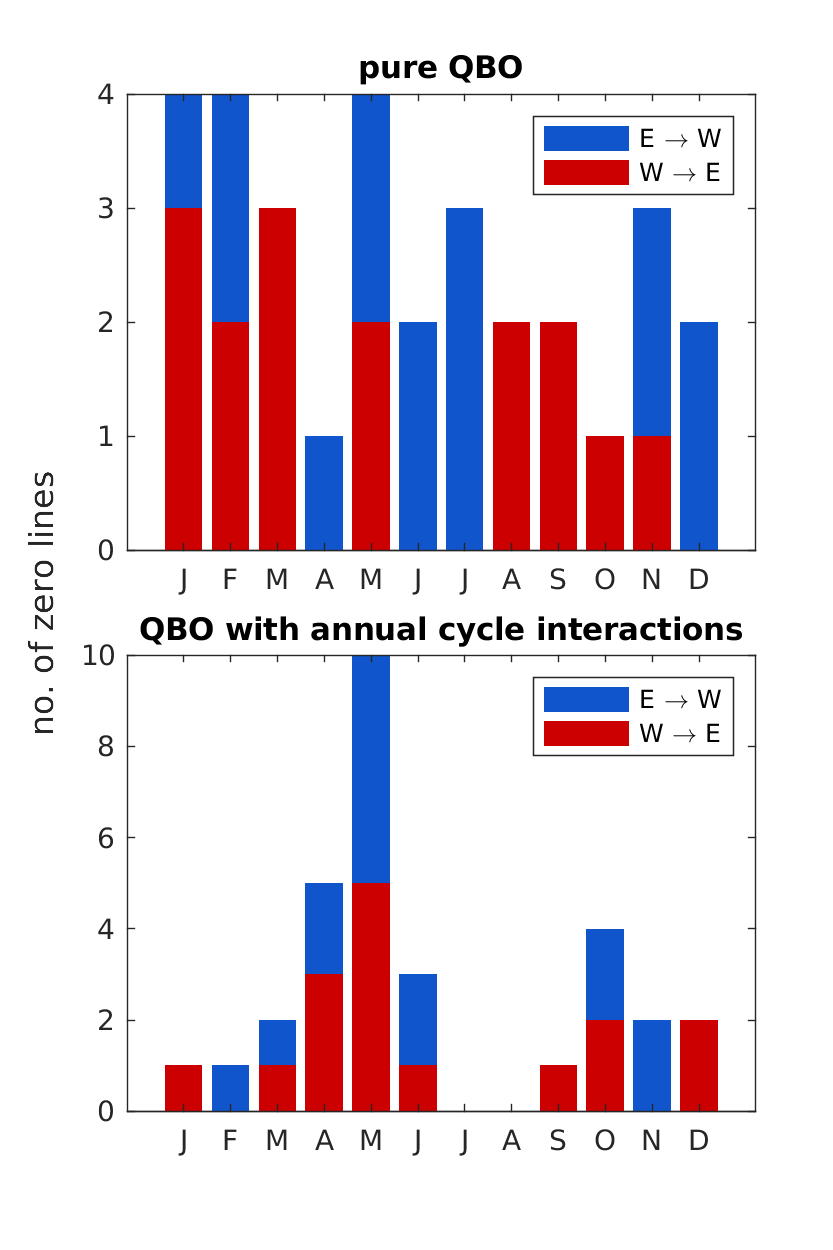}
    \caption{Histograms of zero-line crossings at 50 hPa in the pure QBO mode (top) and the aggregate mode with nonlinear annual cycle interactions (bottom). Colors denote either an easterly to westerly transition (red) or a westerly to easterly transition (blue). While the histogram counting the pure QBO mode exhibits a uniform pattern zero-crossings, the bottom plot shows preferential transitions in boreal spring and no crossings in July or August.}
    \label{fig:phaselocking}
\end{figure}

\begin{figure}
    \centering
    \includegraphics[width=0.5\textwidth]{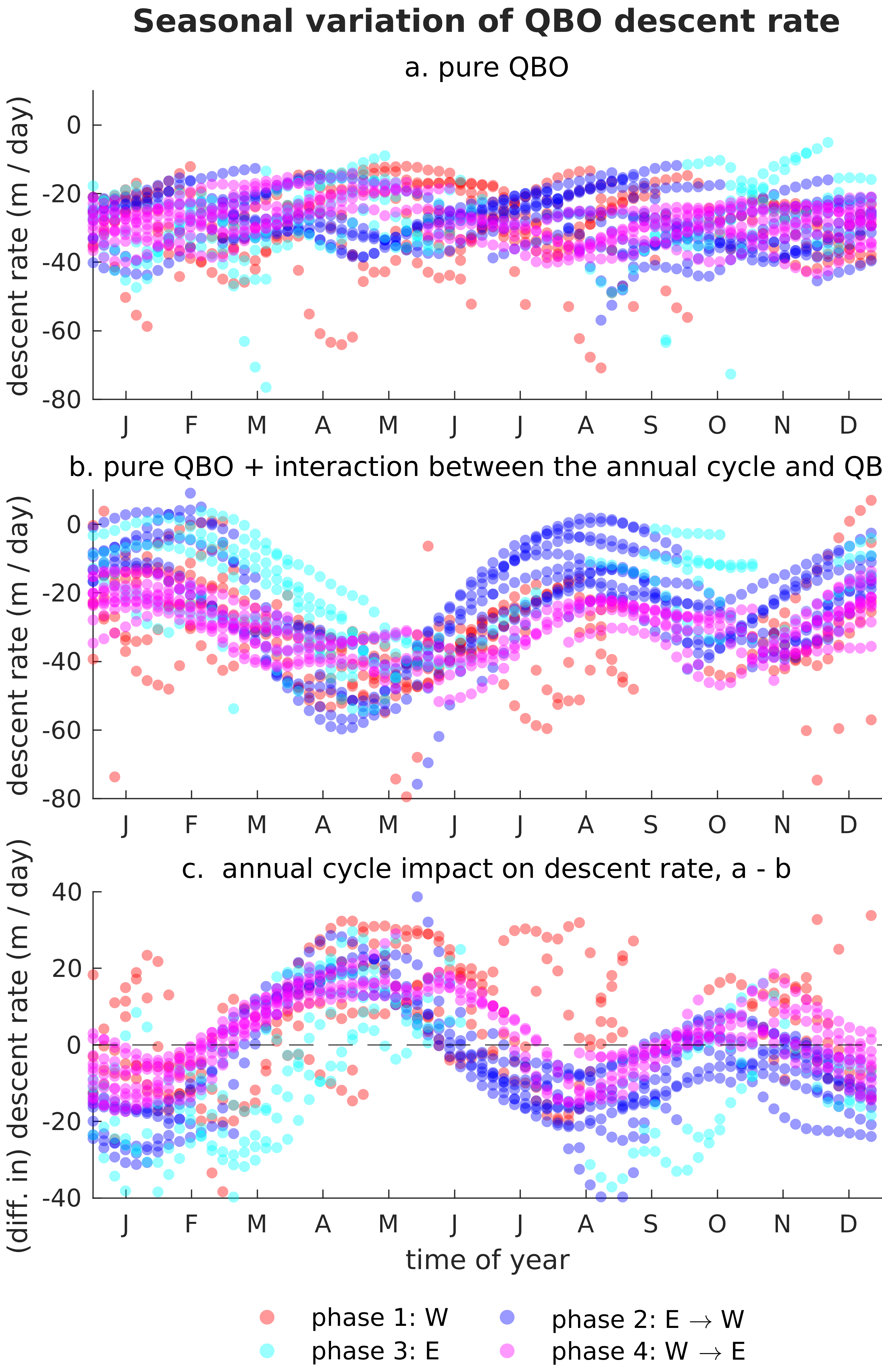}
    \caption{Comparison of QBO zero-line descent rate by time of year between (a) the pure QBO aggregate mode and (b) the QBO with the addition of non-linear QBO-annual cycle interactions. Each scatter point is colored by the phase (as defined by the Koopman eigenfunction metric in figure \ref{fig:phases}) that the QBO was in when the rate was measured. Panel c is the difference between the two above panels, and represents the change in descent rate of the QBO due to the nonlinear interaction between the annual cycle and the QBO. A negative value means that the annual cycle-QBO interactions during this time of year \textit{slow} the descent, rate while a positive difference means that the QBO will descend \textit{faster}. }
    \label{fig:alldr}
\end{figure}

Studies of the influence of the annual cycle on QBO descent rates (specifically the effects of tropical upwelling) have been done, notably using descent rate models, such as in \citet{rajendranDescentRateModels2018}. Previous work was somewhat limited in that conventional methods cannot access a version of the QBO that is not influenced by the annual cycle. As such, an investigation into how the descent rate of the QBO is affected by the seasonal cycle must either reform the problem as seasonal preference of phase onsets, as done in \citep{dunkertonAnnualVariationDeseasonalized1990}, or look at trends in the data and compare to theoretical models. 

% The Koopman decomposition however, allows us to directly study how the descent rate of the QBO (as defined by the descent of the zero-wind line) varies with the annual cycle. 

The onset months of the QBO westerly or easterly phases have been observed to have a seasonal preference and the period of the QBO can vary significantly from the 28-month mean \citep{dunkertonAnnualVariationDeseasonalized1990, hampsonPhaseAlignmentTropical2004}. The perturbation of the QBO by the annual cycle is thought to be the primary reason for the variation in period between cycles, although ENSO and the solar cycle may also contribute. 

The Koopman formulation of the QBO allow us to see the seasonal preference of onset months as a result of nonlinear interactions of the annual cycle. Figure \ref{fig:phaselocking} contains histograms of zero-line crossings at 50 hPa in the pure QBO aggregate Koopman mode and the aggregate mode with additional annual cycle interactions. In the histogram of the aggregate mode that includes both the QBO and annual cycle interactions, we see preferential seasonal transitions from easterly to westerly winds (or vice versa) with a significant peak in spring (May) and secondary peak in fall (October) — similar to those analyzed in previous literature \citep{hampsonPhaseAlignmentTropical2004}. A transition in July or August has never been observed. In contrast, the transition months of our pure QBO mode show little preference in the timing of the onset. We hypothesize that these transition months are drawn from a uniform distribution, i.e., the pure QBO mode has no preferential onset.  We can test this with a $\chi^2$ test, choosing the null hypothesis to be that the QBO transition month is drawn from a uniform distribution; the expectation for the number of QBO transitions in each month is 31/12, given the 31 transitions in the analyzed record, not counting the QBO disruption in 2016. For the pure QBO, we cannot reject the null hypothesis, with $\chi^2 = 5$ and a $p-$value of 0.93. On the other hand, for the aggregate mode that contains the QBO and annual cycle interactions, we find that $\chi^2 = 32.8$ and $p = 5.5 \times 10^{-4}$, meaning that it would be quite unlikely for there to be a uniform preference of QBO transitions. This statistical result is not perfect, as there are a limited number of QBO transitions in the data and each transition is not truly independent of any other. However, the differences in these two aggregate modes strongly indicate that the distribution of QBO transitions is controlled by nonlinear interactions with the annual cycle.

% my multinomial info from here https://www.stat.berkeley.edu/~stark/SticiGui/Text/chiSquare.htm 

% \subsubsection{Computation of zero-crossing descent rates}
To probe the annual variation in QBO further, we compute the descent rate of the QBO by tracking the zero-wind line, that is the descent of the $E \to W$ or $W \to E$ phase transition. One could equivalently track the westerly or easterly maxima, but we found that this introduces more noise (not shown), as the height of the maximum is more sensitive to small perturbations. For a given wind profile, we calculate the location of the mean zero-wind line between $-10^\circ$ and $10^\circ$ N, estimating the exact pressure by interpolating in log-pressure (as recommended in \citet{hersbachERA5GlobalReanalysis2020}). We restrict the zero-line to pressures 10 to 125 hPa (inclusive), to limit issues of defining a descent rate when there are multiple zero lines in a single snapshot (as is often the case when a new phase begins to descend from the upper stratosphere while the other is ending). We compute the descent rate as the derivative computed by a centered difference and then smoothed by a 30-day rolling window. We exclude data from periods where the zero-line ``jumps up,'' i.e., there is a change between descending easterly to westerly winds (or vice versa).

% \subsubsection{Variation in QBO descent rates by the annual cycle}
In figure \ref{fig:alldr}, we plot the zero-crossing descent rates as a function of the time of year. Panel (a) shows the descent rate computed from the pure QBO aggregate mode, where color differentiates the phases of the QBO. The phases correspond in color and definition with those shown in figure \ref{fig:phases}. The systematic undulations of the trajectories reflect changes in descent rate with height, but consistent with the Koopman definition of the pure QBO aggregate mode, there is no dependence of the pure QBO descent rate on the time of year. 

Panel b is the same as panel a, but computed from the aggregate Koopman mode that includes the nonlinear QBO-annual cycle interactions. The third panel (c) is the  \textit{difference} in descent rates between the two above panels and quantifies how much the amount annual cycle-QBO interactions either slow or speed up the descent of the QBO. We observe a marked semiannual modulation of descent rate in the QBO aggregate mode with additional annual effects, both in the raw descent rate (panel b) and more clearly in the difference from the pure QBO descent rate (panel c). This semiannual pattern is noisier for phases $1$ and $3$ (pure westerly or easterly). We expect phases 1 and 3 to be more variable than 2 and 4. Phases 2 and 4 are descending westerly and easterly respectively, where the zero-wind line cleanly descends between 10 and 125 hPa. Phases 1 and 3 contain ``jumps'' of the zero-line between the bottom of the QBO and the top, leading to much of the spread. 

Koopman analysis extracts a semiannual variation in descent rates of order $30 \, \mathrm{m} \, \mathrm{day}^{-1}$, which is comparable in magnitude to the mean descent rate. There is pronounced annual variability, with a stronger peak in May than October. The peaks in May and October are consistent with the enhanced number of QBO transitions during these months. We can similarly explain the few QBO transitions in January and August, as the descent of the zero-line essentially stalls completely during these months. 

Three potential mechanisms for the perturbation of QBO descent rates have been identified by \citet{hampsonPhaseAlignmentTropical2004}: (1) the annual cycle in tropical upwelling caused by the Brewer-Dobson circulation; (2) seasonal variations in the wave forcing that drive the QBO; and (3) seeding from above by the semiannual oscillation (SAO).

\begin{figure*}
    \centering
    \includegraphics[width=1\textwidth]{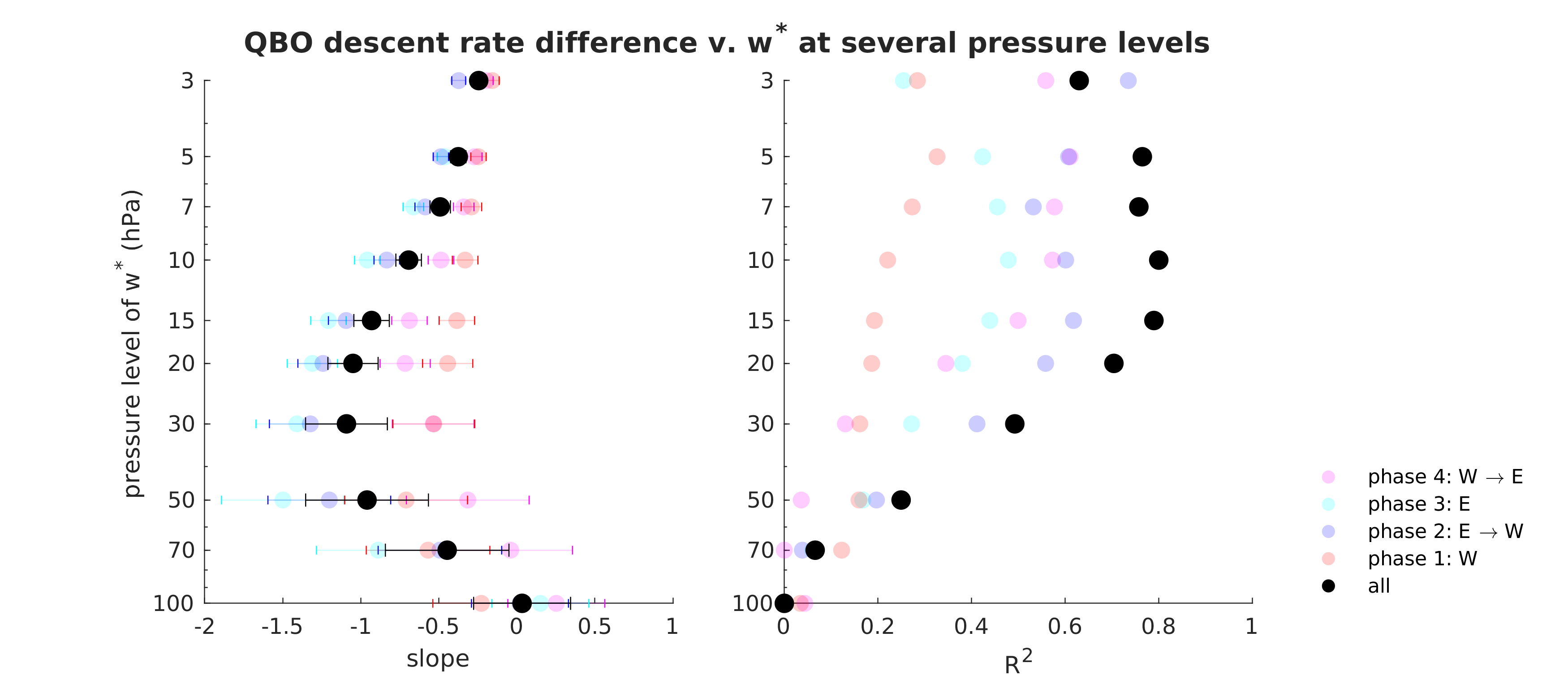}
    \caption{Left: The slope at different pressure levels of the linear fit between descent rate (mean of time of year) vs $w^*$ (mean between $-25^\circ$ and $25^\circ$ N). Error bars denote the 95\% confidence intervals of the slope. Black scatter points correspond to the time of year mean in descent rate over all phases of the QBO, while colored points correspond to each phase. Slope magnitudes peak at 30 hPa. 
    Right: Corresponding $R^2$ (variability explained) values for each of the linear fits in the left figure. The $R^2$ values peak at 15 hPa for most phases. }
    \label{fig:dr_fit}
\end{figure*}

We first investigate to what extent seasonal variation in tropical upwelling is associated with variation in QBO descent rates. Figure \ref{fig:dr_fit} shows the relationship between variation in descent rates and tropical upwelling. We compute the linear relationship between the annual cycle of the transformed Eulerian mean tropical upwelling ($w^*$) as a function of pressure to the mean annual cycle in descent rates. We consider the residual vertical velocity $w^*$ between $3$ and $100 \, \mathrm{hPa}$, as computed by \citet{servaChangesStratosphericDynamics2024}. The black markers correspond to annual mean descent rates regardless of QBO phase, while other colors denote subdivision into one of the QBO phases. 

The left panel shows the regression coefficient and its $95\%$ confidence interval at each pressure level. It indicates how much a $1 \, \mathrm{m} \, \mathrm{day}^{-1}$ change in $w^*$ impacts the descent rate of the QBO. Naively, if changes in the QBO descent rate were the result of a simple perturbation of the advection, we would expect a slope of $-1$. The right panel shows the squared correlation coefficient or $R^2$ of the corresponding linear fits on the left, which suggest what fraction of the variability in mean QBO descent rates can be explained by variation in $w^*$. 

The slope of the linear fits is negligible at $100\, \mathrm{hPa}$, but quickly transitions to nearly around $-1$ from 50 to $15 \, \mathrm{hPa}$. This suggests that the amplitude of annual variations in $w^*$ are consistent with the amplitude of annual variations in descent rate, agreeing with the naive expectation that one can qualitatively explain the descent rate changes simply by changes in advection.  

The change in the correlation coefficient with height is similar to that of the slope. At $100\, \mathrm{hPa}$ the correlation coefficient is essentially zero, suggesting that the annual variation in $w^*$ at the base of the tropical stratosphere has nothing to do with the annual variation of descent rate. As you go higher in the stratosphere, $R^2$ steadily increases with a maximum between $20$ and $5\, \mathrm{hPa}$, where $80 \%$ of the variation in descent rate are consistent with the variation in $w^*$. However, the value of the slope peaks lower than $R^2$, due to the fact that the magnitude of  variability in $w^*$ increases with height in the stratosphere, while the general shape of the seasonal variation remains the same (see the leftmost panel of figure \ref{fig:1ddr}). Another explanation for this vertical structure is that $w^*$ exhibits a more annual cycle in the lower stratosphere (and hence unrelated to the dominant semiannual cycle that is observed) but becomes more decidedly semiannual in the upper stratosphere. As such, $R^2$ increases with height, where the variability of the QBO more strongly matches that of $w^*$.  

In both the combined data and in each QBO phase, $w^*$ potentially accounts for a significant amount of the seasonal variability in QBO descent rates. For the easterly and descending easterly phases (phases 2 and 3), the simple linear relationship appears to work well, though this is diminished for the westerly phases (1 and 4). As different phases of the QBO correspond to different average heights of the zero-crossings and to different dynamics, these changes are not necessarily surprising. We should be cautious not to over interpret these differences between QBO phases, as the subdivision of data decreases the signal-to-noise ratio.

The analysis in figure \ref{fig:dr_fit} also allows us to make predictions about how future changes in tropical upwelling may affect the speed of the QBO descent. For example, consider an increase of $w^*$ by $0.1 \, \mathrm{mm} \, \mathrm{s}^{-1} \approx 9\, \mathrm{m} \, \mathrm{day}^{-1}$, consistent with estimates from quadrupling of $\mathrm{CO}_2$ experiments for several coupled chemistry models \citep{chiodoResponseOzoneLayer2018}. Based on this linear relationship, an increase in upwelling of corresponds to an approximately equal slow down of the QBO descent, translating to an average of $9 \, \mathrm{m} \, \mathrm{day}^{-1}$ decrease in descent rates of the QBO and an increase of its period to 40 months. This prediction assumes there are no changes in other factors that could drive of the QBO descent, such as the wave forcing. Some models do predict an increase in QBO period, but others a decrease \citep{richterResponseQuasiBiennialOscillation2022}.

% could also cite https://essopenarchive.org/doi/full/10.1002/essoar.10512195.1

Correlation analysis cannot establish that $w^*$ causes variations in QBO descent rates, as there could be a third factor that drives both the changes in the QBO and $w^*$. In particular, easonal variations in the wave forcing could influence both $w^*$ and the QBO directly. Similarly, the SAO affects $w^*$ in the upper stratosphere. The wave forcings of the QBO are difficult to diagnose; in particular, gravity waves are not well resolved in the reanalysis. We therefore turn to a simple 1D model to untangle $w^*$ from the wave forcing and the SAO.

\subsection{Interpreting the results with a 1D model of the QBO}

% Three main potential drivers in the annual variability of the QBO descent rates are the SAO, variation in upwelling, and seasonal variations in the wave forcing that drives the QBO. We therefore investigate the QBO descent rate in a 1D QBO model where we can explicitly separate $w^*$ from the wave driving. 

We use a model that is a hybrid of \citet{holtonUpdatedTheoryQuasiBiennial1972} and \citet{plumbInteractionTwoInternal1977}. We solve the following equation for the zonal-wind ($u$):
\begin{equation} \partial_t u + w^* \partial_z u - K \partial_z^2 u = G(z,t) + S(u, z) \end{equation}
where $t$ and $z$ are the time and vertical coordinate, $w^*$ is the (prescribed) vertical advection, $K$ is a constant diffusivity, $G(z,t)$ drives the SAO, and $S(u,z)$ is a monochromatic gravity wave forcing. 

Following \citet{holtonUpdatedTheoryQuasiBiennial1972}, the SAO is prescribed as
\begin{equation}
    G(z,t) = 2G\frac{(z - 28 \, \mathrm{km})}{1000 \, \mathrm{m}} \omega_{\mathrm{SAO}} \sin(2\pi \omega_{\mathrm{SAO}} t)\, \text{ for } z > 28 \, \mathrm{km}
\end{equation} 
where $\omega_{\mathrm{SAO}}$ is the frequency of the SAO. 

The gravity wave forcing $S(u,z)$ assumes two identical gravity waves of opposite phase, as in \citet{plumbInteractionTwoInternal1977} and is detailed in Appendix \ref{appendix:a}. The gravity wave forcing was tuned to give the QBO a 28 month period. 

\begin{figure}
    \centering
    \includegraphics[width=0.5\textwidth]{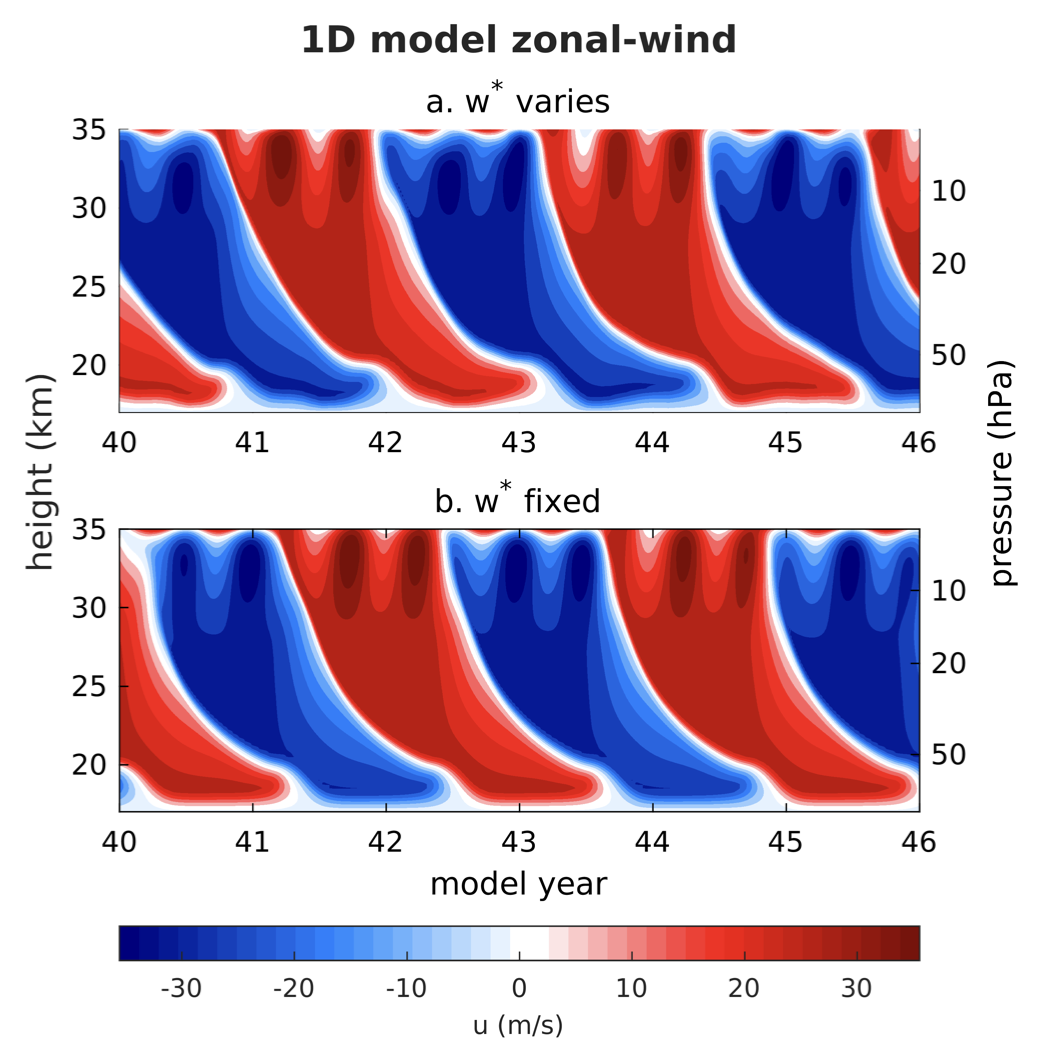}
    \caption{Zonal-wind from the 1D model for model years 40 to 46, both with $G$ (the parameter that sets the SAO strength) equal to 14. Panel a shows zonal-wind from the model where $w^*$ periodically varies annually. Panel b shows zonal-wind from the model where $w^*$ is fixed as the climatology. }
    \label{fig:modelex}
\end{figure}

Each experiment described below reflects a 96 year run, where statistics are excluded for the first 12 years to avoid the influence of initiation. We investigate how prescribing different values of the model parameters affect annual descent rates of the QBO. We define the descent rate of the QBO in the simple model the same way it is defined in the analysis of Koopman modes, by the change in height of the zero-wind line. Portions of the two integrations, with and without annually varying $w^*$ are shown in figure \ref{fig:modelex}.

We first ask if seeding of the SAO perturbs the descent rate of the QBO by augmenting the magnitude of the function $G$.  We tested several values for the magnitude of $G(z,t)$, trying the values $G = 0$, $3.28$, $14$, and $24$ $\mathrm{m} \, \mathrm{s^{-1}}$. The value $G = 3.28 \, \mathrm{m} \, \mathrm{s^{-1}}$ was estimated from reanalysis to match the SAO strength, while $14 \, \mathrm{m} \, \mathrm{s^{-1}}$ is the original magnitude from \citet{holtonUpdatedTheoryQuasiBiennial1972}. We found that the descent rate of the QBO was completely insensitive to the value of G, even when other parameters (such as $w^*$ were varied), indicating that the descent rate only depends on the wave driving and the upwelling in this model.

\begin{figure}
    \centering
    \includegraphics[width=0.5\textwidth]{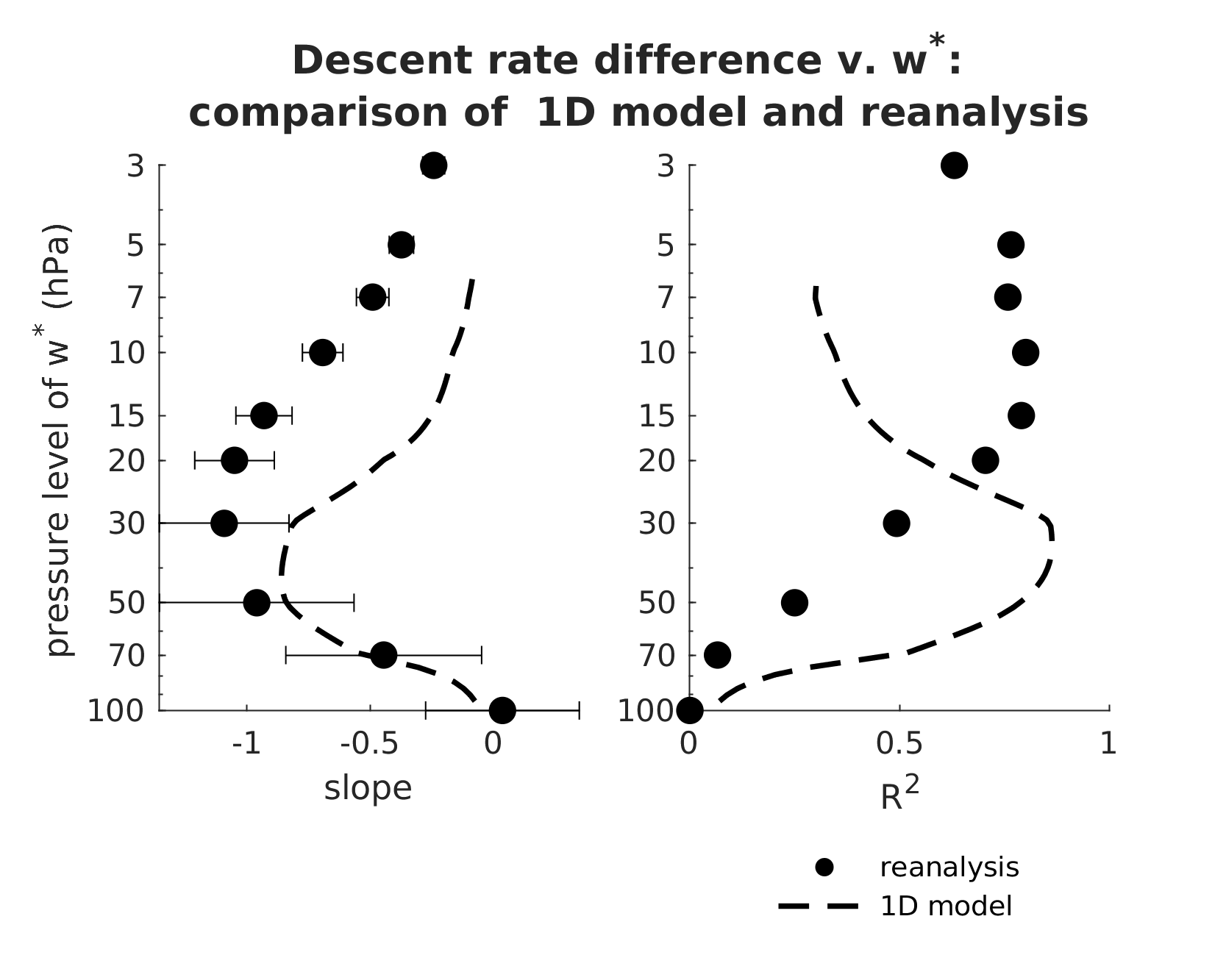}
    \caption{The same as in figure \ref{fig:dr_fit} but for the 1D QBO model (dashed lines) and reanalysis QBO (black circles; same as in figure \ref{fig:dr_fit}, average over all phases). As the 1D model has a top of $35 \, \mathrm{km} \approx 5.8 \, \mathrm{hPa}$, $w^*$ above $5 \, \mathrm{hPa}$ was excluded from the analysis.}
    \label{fig:1dstat}
\end{figure}

\begin{figure}
    \centering
    \includegraphics[width=0.5\textwidth]{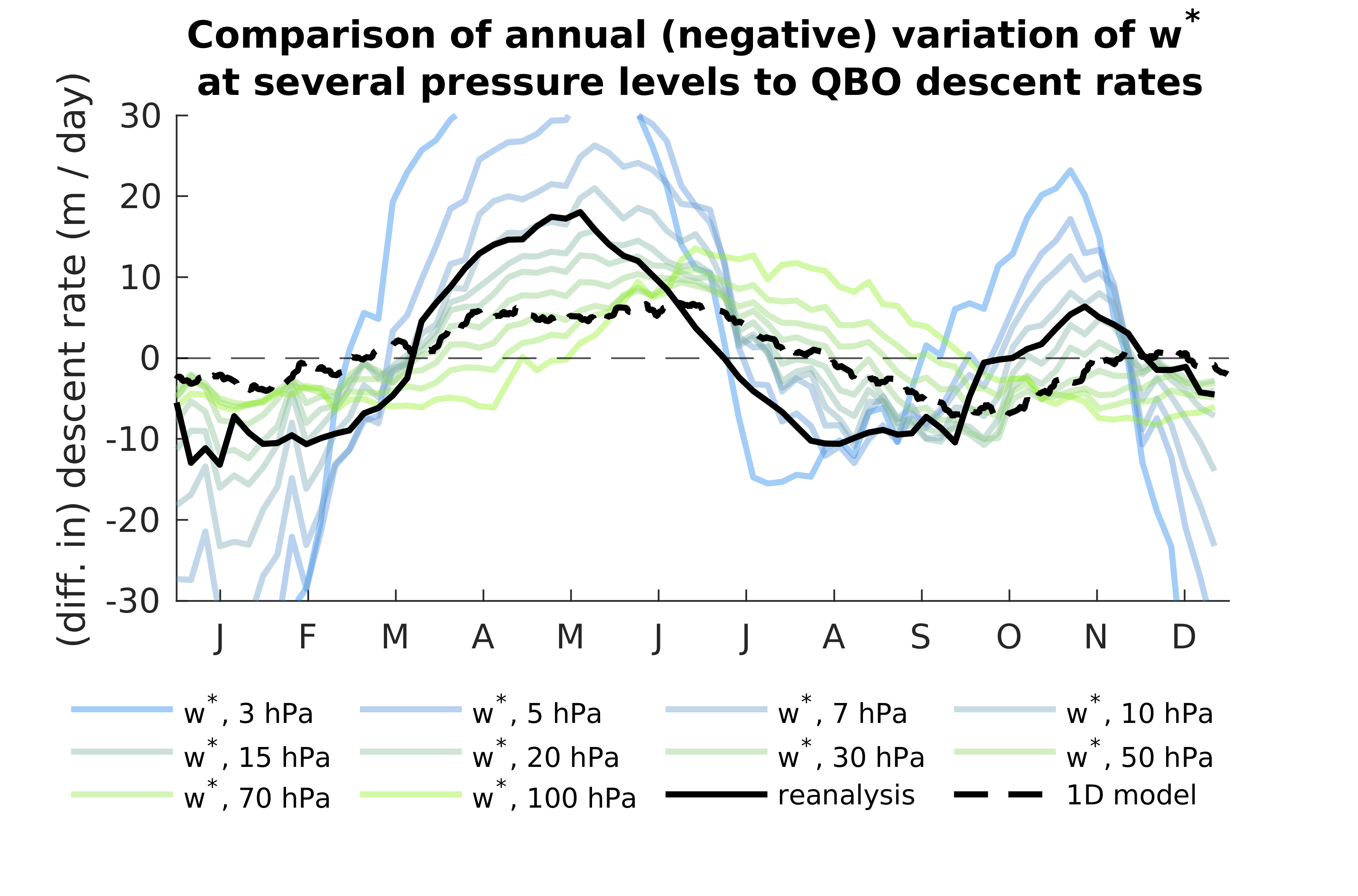}
    \caption{The seasonal cycle in descent rates for reanalysis (solid line; the difference between the pure QBO mode and the mode with added annual-cycle interaction) and the 1D model of the QBO (dashed line; the difference of model runs with a seasonally varying winds and without) with the \textit{negative} deviation from climatology of $w^*$ at several levels. The $w^*$ anomalies are inverted because of their inverse correlation to the descent rate.}
    \label{fig:1ddr}
\end{figure}

We next investigate how much variation the tropical upwelling — which is prescribed independently of the wave forcing — can explain the annual variation of the QBO in the model. We compare model simulations without an annual cycle, where $w^*$ is prescribed to be the climatological mean at each level (analogous to the pure QBO Koopman mode) to simulations with an annual cycle, where prescribed $w^*$ varies with the annual cycle, repeating annually as illustrated in figure \ref{fig:modelex}. We quantify the effect of seasonal upwelling variation on the annual descent rate as the difference between the model run where $w^*$ varies annually and where the model is the climatological mean. 

Figure \ref{fig:1dstat} shows the linear relationship between $w^*$ and annual variation of the 1D model descent rates as a function of pressure. For comparative purposes, we replot the data from the analysis of the Koopman modes in figure \ref{fig:dr_fit}. The slope and correlation coefficient both peak at similar heights (between 60 and 30 hPa). Unlike the observed QBO, the slope that relates the 1D model descent rate to $w^*$ never surpasses -1, but the $R^2$ in the lower stratosphere is higher than in reanalysis. Part of this difference is explained by the extreme regularity of the 1D model compared to the climate system, as well as the longer time period (84 years) in which we compute statistics. Both the slope and correlation coefficient peak lower in the atmosphere for the model than in reanalysis.

To interpret these differences, we compare the 1D model descent rate with that of reanalysis more directly. Figure \ref{fig:1ddr} compares the annual variation in descent rate from the 1D model (black, dotted line) and reanalysis (black, solid line) to the value of $w^*$ at several pressure levels (various colors). The variation in descent rate in reanalysis is about twice that of the 1D model. We also see that while both the reanalysis and the 1D model have a semiannual pattern in annual descent rate, the timing of the peaks are different. 

The fact that the descent rate of the 1D model is more highly correlated with upwelling at lower levels (and less correlated at upper levels) when compared to reanalysis can be understood from the strength of semiannual variation. Higher in the stratosphere, upwelling has a stronger semiannual variation, presumably in part due to the SAO, while lower down (where the 1D model is more sensitive to $w^*$), the annual cycle is more prevalent, causing different maxima in $w^*$. This appears to indicate that the true QBO is more affected by upwelling higher in the stratosphere than in the simple model. This could be due to the fact that stratospheric upwelling itself is more important in our atmosphere, or that the winds higher in the stratosphere are better correlated to another driver of annual changes in the QBO (such as the semiannual oscillation or variations in the wave forcing).

The results from this simple model are consistent with the hypothesis that much of the variability in QBO descent rates are controlled by mean deviations in tropical upwelling, even if the importance of winds at different height varies from the true system. However, upwelling alone cannot reproduce the full seasonal cycle. Annual variation in the wave driving must play an important role in the variation of the descent rate.

\section{Conclusions and future directions}
\label{sec:discussion}
It has been long accepted interactions between the QBO and the annual cycle leads to variation in the downward propagation of the easterly and westerly phases of the oscillation, including their descent rates. Quantitative investigation of these interactions was difficult as conventional data-analysis methods are unable to easily untangle interactions between the QBO and the annual cycle. Koopman formalism is well-suited to this problem, as it is able to separate a periodic element of a system from its nonlinear interactions with other periodic modes of the system. 

We have identified an objective, data-driven index of the QBO that is independent of the annual cycle and performed a decomposition of reanalysis data that separates the effects of the QBO, the annual cycle, and the nonlinear interactions between the two phenomena. This allowed an investigation of the strength of interactions between the QBO and the annual cycle, quantifying to what extent the Holton-Tan effect is driven by nonlinear interactions between the QBO and the annual cycle. We quantify how the annual cycle perturbs the QBO descent rate, comparing it with annual variations in tropical upwelling, with the aid of a classic 1-D model of the QBO.   

The annual variation of the descent rate is large, of amplitude of $30 \, \mathrm{m} \, \mathrm {day}^{-1}$, equivalent to the mean descent rate. As a result, the QBO essentially stalls out from January to February and in July to August (figure \ref{fig:alldr}), explaining the dearth of phase transitions in these months (figure \ref{fig:phaselocking}). Variation in tropical upwelling can account for a large fraction of this variation, particularly in boreal winter. Changes in wave forcing, however, must also play a role, particularly in boreal summer.

Koopman formalism could have further desirable uses for QBO investigation. Applying this method to data that is not zonally averaged would allow for further insight (and possible discovery) of teleconnections between the QBO and higher latitudes that are present on smaller scales than phenomena like the polar vortex. Similar analysis could also be done for other variables of interest, such as temperature or outgoing longwave radiation to probe possible connections between the MJO and QBO \citep{yooModulationBorealWintertime2016}.

Alternatively, we also anticipate that the Koopman decomposition could allow for another avenue of QBO comparison in large climate model evaluations. While advances in gravity wave parameterizations have improved the QBO in comprehensive climate modes, replicating the Holton-Tan effect with observed strength remains a challenge (Garfinkel et al., 2018; Zhang et al., 2019). Using the decomposition as we did for the Holton-Tan effect (\ref{sec:results}\ref{sec:holtontan}) could help to understand where climate models get the Holton-Tan effect wrong: is the QBO itself not well enough resolved, or is it the nonlinear interactions between the QBO and the annual cycle that are missing. 

Finally, the modes found from the Koopman decomposition could also be used for investigation and prediction of other aspects of the climate. Already, these methods have been used successfully to find indices with improved predictability in both ENSO and the MJO when compared to classical methods \citep{wangExtendedrangeStatisticalENSO2019,lintnerIdentificationMaddenJulian2023}.

\clearpage
%%%%%%%%%%%%%%%%%%%%%%%%%%%%%%%%%%%%%%%%%%%%%%%%%%%%%%%%%%%%%%%%%%%%%
% ACKNOWLEDGMENTS
%%%%%%%%%%%%%%%%%%%%%%%%%%%%%%%%%%%%%%%%%%%%%%%%%%%%%%%%%%%%%%%%%%%%%
\acknowledgments
 We acknowledge support from the US National Science Foundation through award OAC-2004572 and the US National Science Foundation Graduate Research Fellowship under Grant DGE-1839302.

%%%%%%%%%%%%%%%%%%%%%%%%%%%%%%%%%%%%%%%%%%%%%%%%%%%%%%%%%%%%%%%%%%%%%
% DATA AVAILABILITY STATEMENT
%%%%%%%%%%%%%%%%%%%%%%%%%%%%%%%%%%%%%%%%%%%%%%%%%%%%%%%%%%%%%%%%%%%%%
% 
\datastatement
The code used to compute the Koopman decomposition can be found on github at: \url{https://github.com/clairevalva/koopmanQBO}. Zonal-wind data from ERA5 can be found at 
\url{https://doi.org/10.24381/cds.bd0915c6} and the $w^*$ from \citet{servaChangesStratosphericDynamics2024} can be found at \url{https://doi.org/10.5281/zenodo.7081436}. The 1D QBO model is implemented in \textit{PyTorch} and can be found at \url{https://github.com/DataWaveProject/qbo1d}.

%%%%%%%%%%%%%%%%%%%%%%%%%%%%%%%%%%%%%%%%%%%%%%%%%%%%%%%%%%%%%%%%%%%%%
% APPENDIXES
%%%%%%%%%%%%%%%%%%%%%%%%%%%%%%%%%%%%%%%%%%%%%%%%%%%%%%%%%%%%%%%%%%%%%
%
%% If only one appendix, use

\appendix
\subsection{Algorithm Overview}
\label{appendix:c}
The algorithm used to compute the Koopman decomposition from data was developed in \citet{dasReproducingKernelHilbert2021}. As noted in the main text, we recommend the appendix and supplement of \citet{froylandSpectralAnalysisClimate2021} and \citet{lintnerIdentificationMaddenJulian2023} for thorough explanations of the algorithms in the climate data context. Here, we give an overview of the computational steps used in this algorithm, and refer  readers to the associated GitHub repository for full parameter and code information: \url{https://github.com/clairevalva/koopmanQBO}.

\paragraph{Delay embed data}
Replace the input data $\mathcal{D}$ with a larger $\hat{\mathcal{D}}$ of size $(N_t - (N_e - 1)) \times (N_d \cdot N_e)$, so that
\begin{equation}
    \hat{\mathcal{D}}_t = (\mathcal{D}_t, \mathcal{D}_{t - 1}, \dots , \mathcal{D}_{t - (N_e - 1)}). \tag{\ref{eq:delay}}
\end{equation}  

\paragraph{Construction of nonlinear basis} We construct nonlinear basis functions $\{\varphi_j\}$ from an eigendecomposition of a kernel matrix $(D)$ constructed from delay embedded data. We define $D_{ij} = k(\mathcal{\hat{D}}_i, \mathcal{\hat{D}}_j)$ where $k$ is a symmetric positive definite kernel function. Then, the basis functions $\varphi_j$ are determined from the following eigendecomposition.
\begin{equation}
    D \varphi_j = \nu_j \varphi_j
\end{equation} This is equivalent to nonlinear Laplacian spectral analysis (NLSA) \citep{giannakisNonlinearLaplacianSpectral2012}. We truncate our basis to have total dimension $N$.

\paragraph{Approximation of Koopman generator in $\{\varphi_j\}$ basis}
Recall the formulation of the Koopman generator $V$:
\begin{equation}
   Vg = \lim_{t \to 0} \frac{K^t g - g}{t}. \tag{\ref{eq:twoa}}
\end{equation}
The application of the approximate operator $\tilde{V}$ acting on a basis function $\varphi_j$ is approximated with a finite difference scheme. Then $\tilde{V}$ (the approximate Koopman generator in the $\varphi_j$) is \textit{symmetrized} to give a unitary operator: $V = (\tilde{V} - \tilde{V}^*) / 2$.

\paragraph{Regularize operator with diffusion} A small amount of diffusion is added to the Koopman generator $V$ for regularization,
\begin{equation}
    W = V - \alpha D,
\end{equation}
where $\alpha$ is a small postive parameter.

\paragraph{Compute eigendecomposition}
The final eigenfunction and eigenvalue pairs $(\omega_j, \zeta_j)$ come from the eigendecomposition of $W$.
\begin{equation}
    W \zeta_j = \omega_j \zeta_j
\end{equation}

\paragraph{Project data on eigenfunctions to create Koopman modes}
Project data $\mathcal{D}$ onto the eigenfunctions $\zeta_j$ to get Koopman modes $M_j$.

% \appendixtitle{1d model of the QBO}
\subsection{1D model of the QBO}
\label{appendix:a}
For our experiments in Section \ref{sec:results}\ref{sec:descentrate} comparing reanalysis and a 1D model, we use a simple model of the QBO that is a hybrid of \citet{holtonUpdatedTheoryQuasiBiennial1972} and \citet{plumbInteractionTwoInternal1977}, as implemented in \textit{PyTorch} by \citet{connellyPytorch1DQBO2022}. We solve the following equation for the zonal-wind ($u$):
\begin{equation} \partial_t u + w^* \partial_z u - K \partial_z^2 u = G(z,t) + S(u, z) \end{equation}
where $t$ and $z$ are the time and vertical coordinate, $w*$ is the (prescribed) vertical advection, $K$ is the diffusivity ($0.3\, \mathrm{m^{2}s^{-1}}$), $G(z,t)$ is the semiannual oscillation, and $S(u,z)$ is the wave forcing.  

The wave forcing term parametrizes the momentum deposition at critical levels as $S(u,z) = \frac{-1}{\rho} \partial_z F(u,z)$ where $\rho(z)$ is the density profile and the wave flux $F(u,z)$ is written as:
\begin{equation}
    F(u,z) = \sum_i A_i \exp(-\int_{zL}^z \frac{\alpha(z) / N}{k_i(u - c_i)^2})
\end{equation}
where $A_i$ and $c_i$ are the wave amplitudes and wave speeds, and $\alpha(z)$ is the wave dissipation due to infrared cooling (see \citet{holtonUpdatedTheoryQuasiBiennial1972}).  Here, we choose a two-wave set up and choose $A_i = \pm1.18e-3$ and $c_i = \pm29$ in order to obtain a 28-month QBO period. 

The semiannual oscillation is prescribed as:
\begin{equation}
    G(z,t) = 2G\frac{(z - 28 \, \mathrm{km})}{1000 \, \mathrm{m}} \omega_{\mathrm{SAO}} \sin(2\pi \omega_{\mathrm{SAO}} t)\, \text{ for } z > 28 \, \mathrm{km}
\end{equation} 
where $\omega_{\mathrm{SAO}}$ is the frequency of the SAO. We found that results of descent rates of the models were insensitive to the amplitude ($G$) of the SAO term. 
% All results shown in this work have $G = 3.28$ (estimated from reanalysis) but are similar to model runs with $G = 0, 14, $ or $24$.   

We choose the model to have a domain top of $z_T = 35\,\mathrm{km}$ and bottom of $z_L = 17\,\mathrm{km}$ with a grid spacing of $250\,\mathrm{m}$. The model is run for 96 years including a 12-year spin-up period with a time step of 12 hours. The upwelling $w^*$ is prescribed to either be the mean at each level  or to be periodic over the year at each level. Log pressure interpolation was used to match grid spacing.

\subsection{Alternative Holton-Tan figure}
\label{appendix:b}
\begin{figure*}
    \noindent\includegraphics[width=0.9\textwidth]{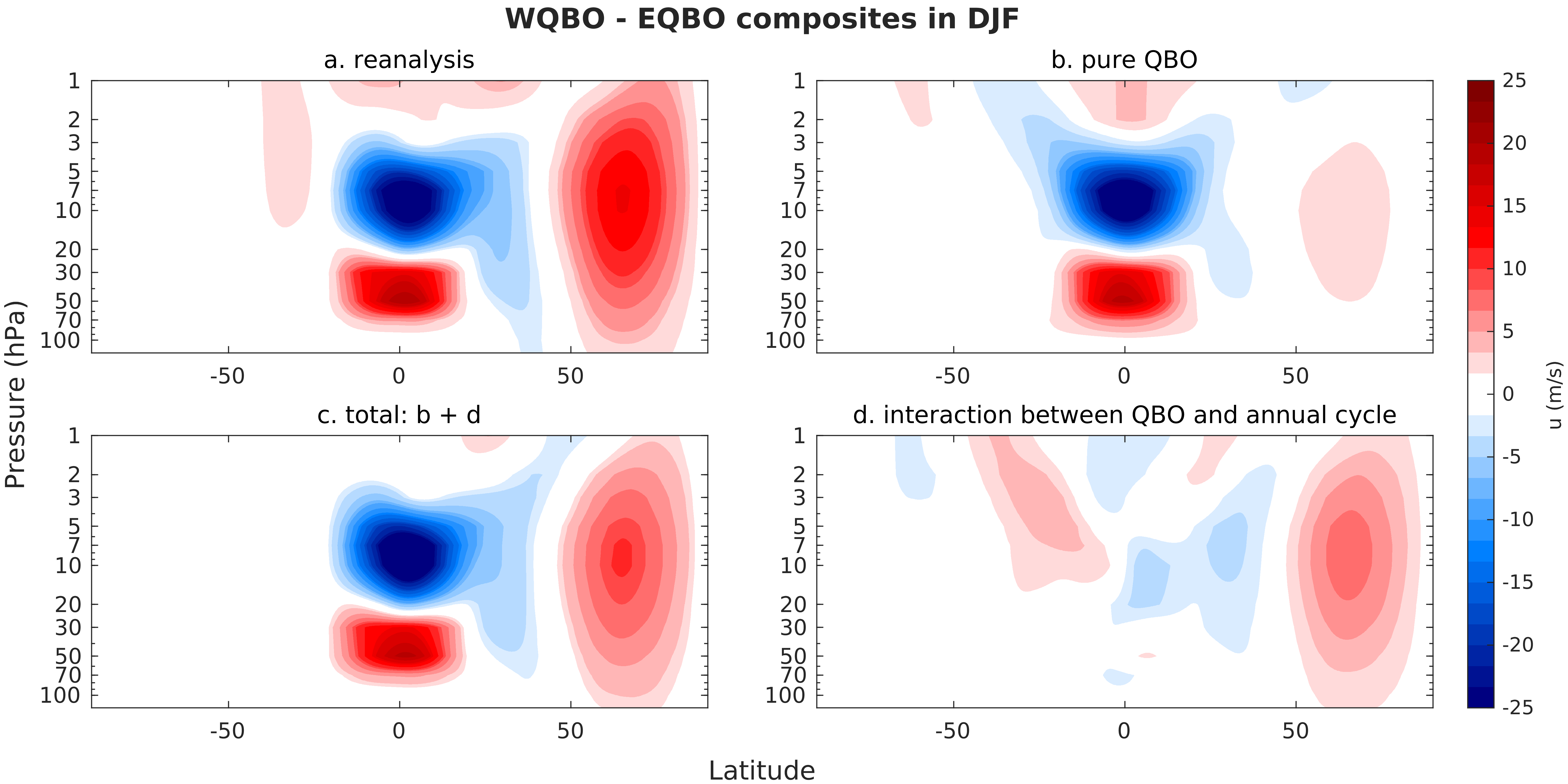}
    \caption{Composites of zonal-mean zonal-wind during boreal winter (DJF) of WQBO-EQBO as measured by the winds of Koopman index composite mode 50 hPa. Panel (a) is reanalysis — the typical composite shown for the Holton-Tan effect. Panels (b) and panel (d) are compositions of the pure QBO and QBO-annual cycle interaction modes. Panel (c) is the sum of (b) and (d). In the case of a perfect reconstruction from the Koopman modes, we would expect (c) to be the same as (a); differences between these panels are on the order of 3 $\mathrm{m} \, \mathrm{s}^{-1}$.}
    \label{fig:HT_alt}
\end{figure*}

The figure \ref{fig:HT_alt} is a variant of figure \ref{fig:HT}, where we determine if each DJF is a WQBO or EQBO season based on the zonal-winds of the pure QBO Koopman aggregate mode. This is opposed to defining based on the unfiltered zonal-winds of reanalysis as in the figure in the main text. The two figures are similar, but not identical, as the pure index flags subtly different winters into the composite.

%%% Appendix section numbering (note, skip \section and begin with \subsection)

%%%%%%%%%%%%%%%%%%%%%%%%%%%%%%%%%%%%%%%%%%%%%%%%%%%%%%%%%%%%%%%%%%%%%
% REFERENCES
%%%%%%%%%%%%%%%%%%%%%%%%%%%%%%%%%%%%%%%%%%%%%%%%%%%%%%%%%%%%%%%%%%%%%
% \section*{References}

\end{document}